\documentclass{aa}
\usepackage{graphics}
\usepackage{times}

\newcommand{\logt}{\mbox{$\log(t/{\rm yr})$}}
\newcommand{\Msun}{\mbox{$M_{\odot}$}}

\newcommand{\sub}[1]{\mbox{$_{\rm #1}$}}

\newcommand{\Teff}{\mbox{$T\sub{eff}$}}
\newcommand{\logTe}{\mbox{$\log T\sub{eff}$}}
\newcommand{\logL}{\mbox{$\log(L/L_{\odot})$}}
\newcommand{\diff}{\mbox{d}}

\begin{document}

	\thesaurus{20(08.05.3, 08.09.3, 08.08.1, 08.12.2) }

        \title{Zero-metallicity stars I. Evolution at constant mass}

	\author{Paola Marigo$^1$, L\'eo Girardi$^1$, Cesare Chiosi$^1$, 
Peter R. Wood$^2$}
	\institute{$^1$ Dipartimento di Astronomia, Universit\`a di Padova,
	Vicolo dell'Osservatorio 2, 35122 Padova, Italia \\
	$^2$ Mount Stromlo and Siding Spring Observatories, Australian National
	University, Private Bag, Weston Creek PO, ACT 2611, Australia}

	\titlerunning{Zero-metallicity stars}
	\authorrunning{P. Marigo et al.}

	\offprints{P. Marigo \\ e-mail: marigo@pd.astro.it}

	\date{}

	\maketitle

	\begin{abstract}
We present extensive evolutionary models of stars 
with initial zero-metallicity, covering a large {range} of initial masses
(i.e. $0.7 M_{\odot} \le M \le 100 M_{\odot}$). Calculations are carried
out at constant mass, with updated input physics, and applying an 
overshooting scheme to convective boundaries. The nuclear network 
includes all the important reactions of the p-p chain,
CNO-cycle and $\alpha$-captures, and is solved by means 
of a suitable semi-implicit method. The evolution is followed up to the 
thermally pulsing AGB in the case of low- and intermediate-mass stars,
or to the onset of carbon burning in massive stars.

The main evolutionary features of these models are discussed,
also in comparison with models of non-zero metallicity. Among
several interesting aspects, particular 
attention has been {paid} to describe: i) the first synthesis of 
$^{12}$C inside the stars, that may suddenly trigger the CNO-cycle 
causing particular evolutionary features; ii) the pollution of 
the stellar surface by the dredge-up events, that are effective 
only within particular mass ranges; iii) the mass limits which conventionally 
define the classes of low-, intermediate-, and high-mass stars on the
basis of common evolutionary properties, 
including the upper mass limit for the achievement 
of super-Eddington luminosities before C-ignition in the high-mass regime; 
and iv) the expected pulsational properties of zero-metallicity stars. 

All relevant information referring to the evolutionary tracks 
and isochrones is made available in computer-readable 
format.


	\end{abstract}  

\section{Introduction}
\label{sec_intro}
The standard Big Bang Nucleosynthesis (e.g. Olive 1999) predicts 
that essentially no elements heavier than $\rm ^{7}Li$ should exist 
in the gas mixture in which the first stars are generated. 
Only in the framework of inhomogeneous Big Bang
nucleosynthesis (see Jedamzik 2000 and references therein) 
we can expect some non-negligible primordial metal production.

The stars with negligible, if not zero, initial metal abundance
-- named population III
objects (Pop-III for short) -- are expected to have structure and 
evolution distinct from those in which traces of heavier elements, 
like those of the CNO group, are present.

The major difference {between zero-metal stars and those} of 
normal metal content (even down to very low values of $Z$) lies in the 
mechanism of nuclear energy generation. In fact, owing to the lack of 
CNO nuclei, the pre-main sequence gravitational contraction cannot 
stop until the central temperature and density
are high enough to allow the p-p chain to provide the energy budget. 
Since the p-p chain is a poor thermostat as compared to the CNO-cycle,
very high temperatures can be reached in the central regions. This 
may eventually lead (depending on the mass of the star) 
to the first synthesis of primary carbon through the 3-$\alpha$ 
reaction, while the star is still on the main sequence.
As a consequence, the CNO-cycle is activated  
at the very high temperatures characterising the 3-$\alpha$ process, 
possibly causing a dramatic change
in the dominant energy source (from p-p chain to CNO-cycle), hence  
affecting the stellar structure amid the main sequence evolution. 
{As demonstrated below}, similar changes in the 
evolutionary behaviour may also occur at later stages, whenever carbon 
is first produced (or transported) into different regions of the star.

The threshold abundance of $\rm ^{12}C$ at which the distinct behaviour 
of zero metal stars appears is as low as $Z \approx X_{\rm C} 
\sim 10^{-10}$ -- $10^{-9}$ (see Cassisi \& Castellani 1993). 
Starting from these values and higher, the efficiency of
CNO-cycle is sufficient to recover the standard behaviour.

Needless to say, the distinct evolutionary behaviour of these stars
may also imply a very distinct nucleosynthesis and chemical pollution
of the interstellar medium, if compared to stars with initial 
$Z>10^{-10}$. Also their properties in the HR-diagram should be quite
peculiar. Both aspects are particularly important in the modelling
of 
	\begin{enumerate}
	\item
the initial chemical enrichment of galaxies, when the first metals 
and additional light elements have been made available to the second 
generation of stars, and probably also to the intergalactic medium;
	\item
the initial spectrophotometric evolution of galaxies, when the 
first ionizing photons of stellar origin probably contributed to 
the reionization of the Universe;
	\item
the observational characteristics that the remaining low-mass
Pop-III stars (if any were formed) would present today, in order to
guide observational searches for these objects.
	\end{enumerate}

In this context, the present paper 
is aimed to provide the necessary tools for the investigation 
of the above-mentioned questions. 
We start by presenting an extensive and homogeneous set of 
evolutionary tracks for Pop-III stars, covering the complete 
mass range from 0.7 to 100$\,M_{\odot}$. 
All of the models are computed with convective
overshooting and under the assumption of constant-mass evolution.
A subsequent paper will deal with the case of more massive stars
($M > 100\,M_{\odot}$), in which stellar winds had likely been 
driven by a state of pulsation instability.

The plan of the paper is as follows.
After briefly reviewing the available literature 
on evolutionary models of primordial stars (Sect.~\ref{sub_rev}),
the main assumptions of present calculations are described in
Sect.~\ref{sec_nuclei} concerning the nucleosynthesis, 
and Sect.~\ref{sec_inp} for the other physics prescriptions.
Then, the presentation and discussion of the results start {with}
Sect.~\ref{sec_general} -- {dealing} with general evolutionary aspects
(e.g. the onset of the 3-$\alpha$ reaction and the activation
of the CNO cycle, lifetimes, etc.) -- and proceed through
Sects.~ \ref{sec_low} and \ref{sec_mas}, {which} focus on particular 
issues of low- and intermediate-/high- mass models, 
respectively (e.g. the AGB phase, the Eddington critical luminosity).
Our predictions for possible changes in the surface chemical
composition are given in Sect.~\ref{sec_chem}.
Finally, Sect.~\ref{sec_strip} is dedicated to {an analysis of} 
the expected pulsational
properties of $Z=0$ stars with different masses and evolutionary stages,
{to investigate} whether pulsation might be a discriminating 
feature in observational searches of Pop-III objects.
Appendices describe the tables of the $Z=0$ evolutionary tracks 
and related  isochrones available in electronic format.

\section {A quick review of past work on Pop-III stars }
\label{sub_rev}
The mass-scale of Pop-III stars is still a matter of debate.
The once widely accepted idea that only massive, very massive or even
super-massive stars could form in the primordial
medium has long limited the study of Pop-III stars to objects more
massive than say $10\,M_{\odot}$, possibly in the range $10^2$ to 
$10^5\, M_{\odot}$ (Carr et al. 1982). 
Similar suggestions were made by Hutchins (1976), Silk (1977) and
Tohline (1980) {whose work} indicated a minimum mass of 
$150-200\, M_{\odot}$.

However, over the years,
interest has turned towards more ``normal'' stars. 
In fact, various fragmentation models of primordial gas have suggested that 
the first stars might have formed with low/intermediate masses, e.g.
Yoshi \& Saio (1986) find the peak of the mass function at roughly 
$4-10 \, M_{\odot}$;  Nakamura \& Umemura (1999) find a the typical 
mass of $3 \, M_{\odot}$, which may grow to $16 \, M_{\odot}$ by accretion; 
and Nakamura \& Umemura (2000a) find a lower mass-limit extending 
down to $1 \, M_{\odot}$. With more detailed and extended 
calculations for the collapse and fragmentation of primordial gas 
clouds, Nakamura \& Umemura (2000b) suggest that the
initial mass function of Pop-III stars may
be bimodal with peaks at about $1 \, M_\odot$ and 
$100 \, M_\odot$, the relative heights of these peaks being a 
function of the collapse epoch. 

The provisional conclusion we may get from all this is that
the mass peak of Pop-III stars is likely greater than say $1 \, M_{\odot}$ and 
possibly in the range
1 to  $10\, M_{\odot}$. However, both lower and much  higher values 
cannot be firmly excluded. 

Ezer \& Cameron (1971) first studied models of Pop-III stars in 
the mass range $5$ to $200 \, M_{\odot}$ from the 
pre-main sequence {until} core H-exhaustion.
In a subsequent paper Ezer (1972) investigated the evolution of a 
$3 \, M_{\odot}$ star up to the stage of core He-exhaustion.
Cary (1974) and Castellani \& Paolicchi (1975) computed 
zero-age-main-sequence (ZAMS)
models in the mass range 2 to $20 \, M_{\odot}$ and 1 to 
$8 \, M_{\odot}$, respectively.

Woosley \& Weaver (1981) followed the evolution at constant mass
of a $200 \, M_{\odot}$  star {until} the supernova explosion stage, thus 
allowing the first estimate of the contamination power of this type of star.
Very massive Pop-III stars were first investigated by Bond et al. 
(1982, 1983).
El Eid et al. (1983) and Ober et al. (1983) explored the  mass range 
from 80 to $500 \, M_{\odot}$, including
mass loss by stellar winds up to the stage of supernova explosion or
collapse to a black hole. Similar studies were undertaken by Klapp (1983,
1984) for very massive stars in the range 500 to 
$10^4\, M_{\odot}$. These papers 
made it possible to get an estimate of the chemical yields both in the 
wind and explosive stages.
Forieri (1982) studied the evolution of massive
Pop-III stars in mass range $10$ to $100 \, M_{\odot}$ up to the central 
He-exhaustion stage. These models were evolved at constant mass
but included convective overshoot.
Castellani et al. (1983) presented models for $10$, $15$ and 
$25 \, M_{\odot}$ stars from ZAMS to C-ignition.

The advanced phases of low-mass stars remained almost unexplored 
till the early study 
by D'Antona  (1982) of a $1 \, M_{\odot}$ star up to the stage of core 
He-flash, where for the first time the much less extended RGB was noticed
(see also Castellani \& Paolicchi 1975 for an earlier suggestion).
Chieffi \& Tornamb\'e (1984) investigated the evolution of a 
$5 \, M_{\odot}$ object, noticing the absence of thermal instabilities 
during the double-shell (AGB) phase.  
Fujimoto et al. (1984) pointed out that threshold values for 
the core mass and CNO abundances exist for the occurrence of 
He-shell flashes during the AGB phase. 
Cassisi \& Castellani (1993) presented grids of models with masses in the 
range  $0.7 \leq  M \leq 15 M_{\odot}$ and  metallicity from 
$Z=10^{-4}$ down to $Z=10^{-10}$. The models were followed
from ZAMS till He-flash, final cooling of the CO core
or C-ignition as appropriate. Cassisi et al. (1996) analysed 
the RR Lyrae properties and the unusually strong He-shell flashes 
of low-mass stars with very low initial metal content 
($-10 \le \log Z \le -5$).

Many recent/ongoing works on zero-metallicity stars have been recently
presented in the symposium proceedings edited by Weiss et al. (2000a),
to which the reader should refer for a comprehensive picture
of the present {state of understanding of these objects}.
Particularly impressive are the studies being carried out by 
Heger et al. (2000) and Umeda et al. (2000), where massive 
zero-metallicity stars are made to evolve up to the pre-supernova stage 
and supernova explosion, respectively. These studies describe in
detail the nucleosynthesis of heavy elements occurring in these stars.

In the work by Weiss et al. (2000b) the authors investigate the role of
diffusion and external pollution in the evolution of metal-free 
low-mass stars, finding that no sizable effects occur (the accreted
metals cannot reach the central H-burning regions, 
which would have otherwise affected the structure of zero-metallicity stars).  
Using an extended nuclear network, they also demonstrate 
that hot p-p chains are less effective than the 3-$\alpha$ process in 
producing carbon inside a zero-metallicity star.

{Finally, we should mention the work by Schlattl et al. (2001),
who investigate the possible changes in the envelope chemical composition
as the result of convective dredge-up during the He-flash of Pop-III RGB stars
(see also Fujimoto et al. 2000).}

\section{Nuclear network and nucleosynthesis}
\label{sec_nuclei}
In general, the evaluation of {temporal} changes in the chemical abundances due to
nuclear reactions requires the integration of the system of equations below: 
\begin{equation}
\frac {d}{dt} Y_i = -[ij]\:\: Y_i Y_j + [rs]\:\: Y_r Y_s, 
\:\:\:\:\:\:i = 1,...N_{\rm el} 
\label{abbeq}
\end{equation}
where $N_{\rm el}$ is the number of elemental species involved; 
$Y_i = X_i/A_i$, $X_i$, and $A_i$ denote the abundance by number (mole
g$^{-1}$), the abundance by mass, and the atomic mass of the elemental species
$i$, respectively. 
In the right-hand side of the equation (\ref{abbeq}), $[ij]$
stands for the rate of the generic reaction which converts the element $i$ into
another nucleus because of the interaction with the element $j$, whereas $[rs]$
is the rate of the generic reaction transforming nuclei $s$ and $r$ into the
element $i$. 

In this work the adopted nuclear network consists of the
sets of reactions for the pp1-chain
\begin{equation}
\begin{array}{llll}
{\rm\, ^1H\, +\, ^1H\, \longrightarrow\, ^{2}H + \beta^{+} + \nu} \\
{\rm\, ^2H\, +\, ^1H\, \longrightarrow\, ^{3}He + \gamma} \\
{\rm\, ^3He\, +\, ^3He\, \longrightarrow\, ^{4}He + 2 ^1H} \\
{\rm\, ^3He\, +\, ^4He\, \longrightarrow\, ^{7}Be + \gamma}, 
\end{array}
\label{pp1}
\end{equation}
for the CNO tri-cycle
\begin{equation}
\begin{array}{lllllllllll}
{\rm\, ^{12}C\, +\, ^1H\, \longrightarrow\, ^{13}C + \gamma + \beta^{+} + \nu} \\
{\rm\, ^{13}C\, +\, ^1H\, \longrightarrow\, ^{14}N + \gamma} \\
{\rm\, ^{14}N\, +\, ^1H\, \longrightarrow\, ^{15}N + \gamma + \beta^{+} + \nu} \\
{\rm\, ^{15}N\, +\, ^1H\, \longrightarrow\, ^{4}He\, +\, ^{12}C} \\
{\rm\, ^{15}N\, +\, ^1H\, \longrightarrow\, ^{16}O + \gamma} \\
{\rm\, ^{16}O\, +\, ^1H\, \longrightarrow\, ^{17}O + \gamma + \beta^{+} + \nu} \\
{\rm\, ^{17}O\, +\, ^1H\, \longrightarrow\, ^{4}He\, +\, ^{14}N} \\
{\rm\, ^{17}O\, +\, ^1H\, \longrightarrow\, ^{18}O + \gamma + \beta^{+} + \nu} \\
{\rm\, ^{18}O\, +\, ^1H\, \longrightarrow\, ^{4}He\, +\, ^{15}N} \\
{\rm\, ^{18}O\, +\, ^1H\, \longrightarrow\, ^{19}F + \gamma} \\
{\rm\, ^{19}F\, +\, ^1H\, \longrightarrow\, ^{4}He\, +\, ^{16}O}, 
\end{array}
\label{cno}
\end{equation}
and the most important $\alpha$-capture reactions
\begin{equation}
\begin{array}{llllllllll}
{\rm\, ^{4}He\, +\, 2 ^4He\, \longrightarrow\, ^{12}C + \gamma} \\
{\rm\, ^{12}C\, +\,  ^4He\, \longrightarrow\, ^{16}O + \gamma} \\
{\rm\, ^{14}N\, +\, ^4He\, \longrightarrow\, ^{18}F + \gamma} \\
{\rm\, ^{16}O\, +\, ^4He\, \longrightarrow\, ^{20}Ne + \gamma} \\
{\rm\, ^{14}N\, +\, ^4He\, \longrightarrow\, ^{18}F + \gamma} \\
{\rm\, ^{18}O\, +\, ^4He\, \longrightarrow\, ^{22}Ne + \gamma} \\
{\rm\, ^{14}N\, +\, ^4He\, \longrightarrow\, ^{18}F + \gamma} \\
{\rm\, ^{22}Ne\, +\, ^4He\, \longrightarrow\, ^{25}Mg + n} \\
{\rm\, ^{14}N\, +\, ^4He\, \longrightarrow\, ^{18}F + \gamma} \\
{\rm\, ^{22}Ne\, +\, ^4He\, \longrightarrow\, ^{26}Mg + \gamma} \\
\end{array}
\label{alpha}
\end{equation}

The abundance equations (\ref{abbeq}) 
are solved with the aid of 
a semi-implicit extrapolation
scheme (Bader \& Deuflhard 1983), 
which combines all the considered H- and He-burning reactions
(Eqs.~\ref{pp1}, \ref{cno}, and \ref{alpha}) 
without any assumption for nuclear equilibria.
Such procedure is necessary to follow the nucleosynthesis 
occurring in stars
with original zero (or extremely low) metal content,
for various reasons.

The simultaneous consideration 
of H- and He-burning reactions is required by the fact that 
 in these stars H-burning may occur at  temperatures and densities
high enough to allow the synthesis, via $\alpha$-captures, of the 
missing catalysts (i.e. $^{12}$C and $^{16}$O)
for the activation of the CNO cycle. 

The choice of a semi-implicit method is motivated by 
the search for a convenient  compromise between
the higher accuracy typical of the explicit scheme, and the better stability 
of the solution guaranteed by the implicit scheme.
 
We recall that,
from a technical point 
of view, in implicit/explicit  methods the increments of      
the dependent variables over a given
integration step are calculated from the derivatives
evaluated at the new/old location of the independent variable.
The net effect is that implicit methods are stable for large 
integration steps, but with a certain loss of accuracy in following the
solution towards the equilibrium. The reverse situation occurs
for explicit methods.

Actually, implicit methods are required when dealing with stiff 
differential equations, in which the changes of the dependent variables 
involve quite different scales of the independent variable.
This feature matches the case of the abundance equations, as different 
elemental species  are characterised by very different 
nuclear lifetimes.
For instance, at temperatures near 25 $\times \, 10^6$ K, 
the e-folding time for
$^{15}$N is on the order of years, whereas for $^{14}$N it 
is on the order of $10^{5}$ years.
In general, 
implicit methods converge to the true equilibrium solutions
of the nuclear network for relatively large time steps, 
whereas purely explicit methods would require
extremely short time steps to both find the solution and 
maintain stability. Alternatively, in order to make the  
explicit scheme less time consuming 
this may be coupled with external assumptions 
on nuclear equilibria (usually made for the 
CNO nuclei). This possible choice is however 
quite risky, specially in the case of zero-metallicity
stars in which the total abundance 
by number of the CNO elements during H-burning 
is not constant, but rather increasing 
thanks to the contribution from He-burning.

\section{General physical input of the stellar models } 
\label{sec_inp}

\paragraph{Chemical composition.}
The adopted initial composition consists of a mixture 
of just hydrogen and helium, with mass fractions of
$X=0.77$ and $Y=0.23$, respectively. 
The adopted $Y$ value is consistent with most recent determinations 
of the primordial helium content, that typically range from 
0.230 to 0.234 (see e.g.\ Peimbert 1996 for a review; or 
Olive et al.\ 1997 and Viegas et al.\ 2000 for recent determinations).

\paragraph{Opacity.}
The radiative opacities are from the OPAL group
(Rogers \& Iglesias 1992; Iglesias \& Rogers 1993) for temperatures
higher than $\log T = 4.1$, and from Alexander \& Ferguson
(1994) for $\log T \le 4.0$. In the temperature interval 
$4.0 < \log  T  < 4.1$, a
linear interpolation between the opacities derived from both
sources is adopted. 
The conductive opacities of electron-degenerate matter are
from Hubbard \& Lampe (1969). 

\paragraph{Equation of state.}
The equation of state (EOS) for temperatures higher than $10^7$~K is
that of a fully-ionized gas, including electron degeneracy in the way
described by Kippenhahn et al.\ (1965). The effect of Coulomb
interactions between the gas particles at high densities is
introduced following the prescription by Straniero (1988; cf.\
also Girardi et al.\ 1996).

\paragraph{Nuclear reaction rates and neutrino losses.}
The reaction rates are from the compilation of Caughlan \& Fowler
(1988), {except} for $^{17}{\rm O}({\rm
p},\alpha)^{14}{\rm N}$ and $^{17}{\rm O}({\rm p},\gamma)^{18}{\rm
F}$, for which we use the more recent determinations by 
Landr\'e et al.\ (1990). The
uncertain $^{12}$C($\alpha,\gamma$)$^{16}$O rate was set to 1.7 times
the values given by Caughlan \& Fowler (1988), as indicated by the
study of Weaver \& Woosley (1993) on the nucleosynthesis by massive
stars.  The electron screening factors for all reactions are those
from Graboske et al.\ (1973).

The energy losses by pair, plasma, and bremsstrahlung neutrinos, 
are from Haft et al.\ (1994).

\paragraph{External convection and overshooting.}
The energy transport in the outer convection zone is described
according to the mixing-length theory of B\"ohm-Vitense (1958). The
mixing length parameter $\alpha=1.68$ comes from the calibration of
a solar model (Girardi et al.\ 2000).

The extension of convective boundaries is estimated by means of an
algorithm which takes into account overshooting from 
both core and envelope convective zones. The formalism is fully
described in Bressan et al.\ (1981) and Alongi et al.\ (1991).
The main parameter describing
overshooting is its extent $\Lambda_{\rm c}$ across
the border of the convective zone, expressed in units of pressure
scale height. 

We adopt the following prescription for the
parameter $\Lambda_{\rm c}$ as a function of stellar mass:
	\begin{itemize}
	\item
$\Lambda_{\rm c}=0 \,\,\,\,\,\,$ for $M<~1.1$~\Msun,
	\item 
$\Lambda_{\rm c}=0.5 \,\,\,$ for $M\ge~1.1$~\Msun,
	\end{itemize} 
This is for the stages of core hydrogen burning.
During core helium burning (CHeB), the value $\Lambda_{\rm
c}=0.5$ is used for all stellar masses.

Overshooting at the lower boundary of
convective envelopes is also considered. 
The value of $\Lambda_{\rm e}=0.25$ (see Alongi et al.\
1991, for a description of the formalism) is adopted for 
stars with $0.7\le(M/\Msun)<2.0$. For
$M>2.0$~\Msun\ a value of $\Lambda_{\rm e}=0.7$ is assumed.

\paragraph{Mass loss by stellar winds.}

Perhaps  the most uncertain aspect of the evolution of Pop-III
stars is mass loss by stellar winds. Two typical situations 
 can be envisaged: 
(i) Mass loss from  massive stars during their whole evolutionary history, 
and (ii) mass loss during the
RGB and AGB phases of low and intermediate-mass stars.

Of course, the first question is whether 
{our knowledge}  for normal-metallicity stars  
can be extrapolated and safely used for Pop-III stars.
This depends on the physical nature of the mechanism driving mass-loss.
In massive stars  of normal metallicity (no matter whether low or high)
mass loss at high effective temperature is due to radiation pressure on
resonance lines of ions like CIV, NV, SiIV, etc., 
whereas that at low temperatures
is likely caused by radiation pressure on dust grains and/or 
{due to} pulsational
properties (see Chiosi \& Maeder 1986; 
Chiosi et al. 1992;  Maeder \& Conti 1994; Feast 1991;
Kudritzki 1998; for recent reviews). Furthermore, the
body of observational and theoretical information allows us to
establish empirical relations according to which the mass loss rates are
seen to decrease with the metallicity. 
On this {basis} one could perhaps 
argue that as long as the surface layers keep the original composition 
mass loss cannot occur. If and when the surface abundances are 
contaminated by mixing of heavy elements, mass loss by the above 
mechanisms could be activated.
However, the above reasoning would immediately fail if other mechanisms 
(metallicity independent) concur to drive mass loss. 

Even for non-zero metallicities, modelling  
the RGB and AGB phases of low- and intermediate-mass stars 
{is usually problematic}, as we do not have a definitive theoretical
scenario to account for mass loss during
these stages. Moreover, the question arises {as to} whether the 
mass-loss prescriptions normally used for metal-poor globular cluster
giants (e.g. Reimers 1975), or for pulsating AGB stars (e.g.
Vassiliadis \& Wood 1993) can be applied to Pop-III stars.
Anyway, if any of these empirical relations is actually used in 
zero-metallicity models, very little mass is found to be lost 
during the RGB and AGB phases. This  justifies to some extent the 
assumption of evolution at constant mass, that we adopt
in the present paper. 

However, in Sect.~\ref{sec_mas} we explore the possibility that
the more massive and evolved Pop-III stars may become 
gravitationally unbound and hence lose mass. These stars will be
the subject of a forthcoming paper (Marigo et al.
2001, in preparation).

\begin{figure}
\resizebox{\hsize}{!}{\includegraphics{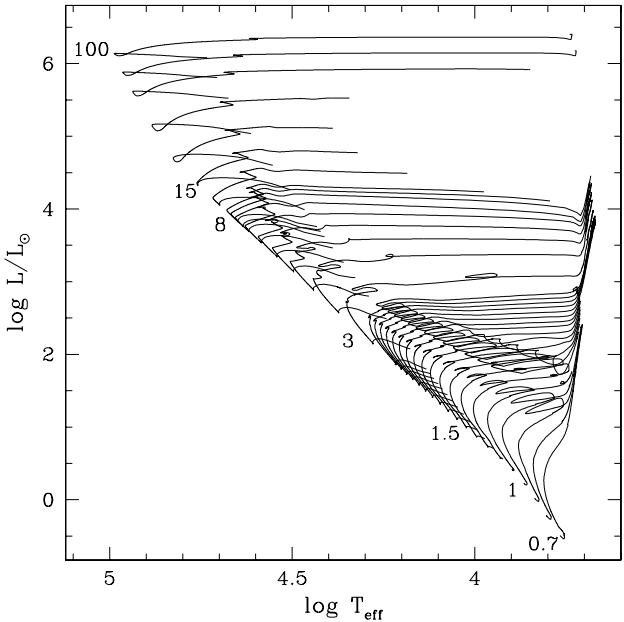}}
\caption{Zero-metallicity stellar 
tracks in the HR diagram, for initial masses in the range 
0.7 $M_{\odot}$ -- 100 $M_{\odot}$. 
Evolutionary calculations are carried out at constant mass.
For the sake of simplicity, the He-burning tracks of
low-mass ($M<1.1\, M_\odot$) are not plotted.
See also Fig.~\protect\ref{fig_hhe}.}
\label{fig_hr}
\end{figure}

\begin{table}
\caption{Nuclear lifetimes as a function of the initial stellar mass.}
\label{tab_time}
\begin{tabular}{rcc}
\noalign{\smallskip}\hline\noalign{\smallskip}
$M/M_{\odot}$ &  $\tau_{\rm H}/{\rm yr}$  &    $\tau_{\rm He}/{\rm yr}$ \\
\noalign{\smallskip}\hline\noalign{\smallskip}
    0.7 &      2.2751 $10^{10}$ &      1.1598 $10^{8}$ \\
    0.8 &      1.3749 $10^{10}$ &      1.1536 $10^{8}$ \\ 
    0.9 &      8.9023 $10^{9}$ &      1.0576 $10^{8}$ \\  
    1.0 &      6.1077 $10^{9}$ &      1.1029 $10^{8}$ \\  
    1.1 &      4.3791 $10^{9}$ &      2.0626 $10^{8}$ \\  
    1.2 &      3.3170 $10^{9}$ &      1.5657 $10^{8}$ \\  
    1.3 &      2.5358 $10^{9}$ &      1.2843 $10^{8}$ \\  
    1.4 &      1.9741 $10^{9}$ &      1.0660 $10^{8}$ \\ 
    1.5 &      1.5810 $10^{9}$ &      9.4372 $10^{7}$ \\  
    1.6 &      1.2830 $10^{9}$ &      7.5237 $10^{7}$ \\  
    1.7 &      1.0510 $10^{9}$ &      6.3585 $10^{7}$ \\  
    1.8 &      8.8129 $10^{8}$ &      6.2271 $10^{7}$ \\  
    1.9 &      7.4387 $10^{8}$ &      4.8001 $10^{7}$ \\  
    2.0 &      6.3454 $10^{8}$ &      4.2688 $10^{7}$ \\  
    2.1 &      5.4827 $10^{8}$ &      3.7334 $10^{7}$ \\  
    2.2 &      4.7794 $10^{8}$ &      3.5347 $10^{7}$\\  
    2.3 &      4.2032 $10^{8}$ &      3.0438 $10^{7}$ \\  
    2.4 &      3.7470 $10^{8}$ &      2.6636 $10^{7}$ \\  
    2.5 &      3.3439 $10^{8}$ &      2.4283 $10^{7}$ \\  
    2.7 &      2.7806 $10^{8}$ &      2.0256 $10^{7}$ \\  
    3.0 &      2.0931 $10^{8}$ &      1.6214 $10^{7}$ \\  
    3.5 &      1.4855 $10^{8}$ &      1.2113 $10^{7}$ \\  
    4.0 &      1.0755 $10^{8}$ &      1.1333 $10^{7}$ \\  
    5.0 &      6.7291 $10^{7}$ &      6.9363 $10^{6}$ \\  
    6.0 &      4.7335 $10^{7}$ &      4.4815 $10^{6}$ \\  
    6.5 &      4.1719 $10^{7}$ &      3.5485 $10^{6}$ \\  
    7.0 &      3.5567 $10^{7}$ &      2.9240 $10^{6}$ \\  
    8.0 &      2.6947 $10^{7}$ &      2.0781 $10^{6}$ \\  
    8.3 &      2.5305 $10^{7}$ &      1.9528 $10^{6}$ \\  
    9.0 &      2.2221 $10^{7}$ &      1.6740 $10^{6}$ \\  
    9.5 &      2.0475 $10^{7}$ &      1.5268 $10^{6}$ \\  
   10.0 &      1.8923 $10^{7}$ &      1.4089 $10^{6}$ \\  
   12.0 &      1.4710 $10^{7}$ &      1.0536 $10^{6}$ \\  
   15.0 &      1.1307 $10^{7}$ &      7.7869 $10^{5}$ \\  
   20.0 &      8.4720 $10^{6}$ &      5.7555 $10^{5}$ \\  
   30.0 &      5.9030 $10^{6}$ &      4.2264 $10^{5}$ \\  
   50.0 &      4.1315 $10^{6}$ &      3.3023 $10^{5}$ \\  
   70.0 &      3.4538 $10^{6}$ &      2.9938 $10^{5}$ \\ 
  100.0 &      2.9509 $10^{6}$ &      2.6841 $10^{5}$ \\ 
\noalign{\smallskip}\hline\noalign{\smallskip}
\end{tabular}
\end{table}

\section{General properties of the stellar tracks}
\label{sec_general}
\subsection{Mass ranges, evolutionary stages, and lifetimes}
\label{sec_massranges}
Figure \ref{fig_hr} shows
the whole set of zero-metallicity tracks at constant 
mass in the $L-T_{\rm eff}$
diagram. The initial masses range from 
0.7 $M_{\odot}$ to 100 $M_{\odot}$.

For the sake of clarity in the presentation of the
results, we adopt the standard convention of sub-grouping the models
as a function of the initial mass, 
by defining two mass limits, namely $M_{\rm HeF}$ and $M_{\rm up}$.
\begin{itemize}
\item {\sl Low-mass stars}, with $M \le  M_{\rm HeF}$, suffer He-ignition
in an electron-degenerate core at the tip of the Red Giant Branch (RGB). 
\item {\sl Intermediate-mass stars}, with $ M_{\rm HeF} < M \le M_{\rm up}$,
ignite He in a non-degenerate core, but later
develop an electron-degenerate C-O core and experience the 
Asymptotic Giant Branch (AGB) phase.
\item {\sl Massive stars}, with $M >  M_{\rm up}$, are able to ignite 
central carbon in non-degenerate conditions and proceed through
the sequence of advanced nuclear burnings.   
\end{itemize}

Our models are evolved from the ZAMS, at constant mass. 
The evolution through the whole H- and He-burning phases is followed 
in detail. The calculations are carried out up to the 
initial stages of the TP-AGB phase in intermediate-
and low-mass models (a small number of thermal
pulses have been followed), 
or the onset of carbon ignition in the helium-exhausted
core for the most massive models. 

For low-mass stars, the evolution is interrupted at the stage of 
He-flash in the electron-degenerate hydrogen-exhausted
core. The evolution is then re-started from a 
Zero Age Horizontal Branch (ZAHB) model having the 
same core mass $M_{\rm c}^{\rm HeF}$ and envelope
chemical composition as the last RGB model (see also Sect.~\ref{ssec_hb}).

Nuclear lifetimes of the H- and He-burning phases as a function 
of the stellar mass are reported in Table \ref{tab_time}. 

Let us now discuss the main features of the models here presented.
It should be recalled that 
many aspects of the evolution of zero-metallicity stars are already
{described} by Cassisi \& Castellani(1993; hereinafter also CC93).
In order to avoid redundant discussion, the present analysis is more
focused on  particular issues and/or open questions which, to our opinion,
still deserve further investigation. 

\begin{figure*}
\begin{minipage}{0.69\textwidth}
\resizebox{\hsize}{!}{\includegraphics{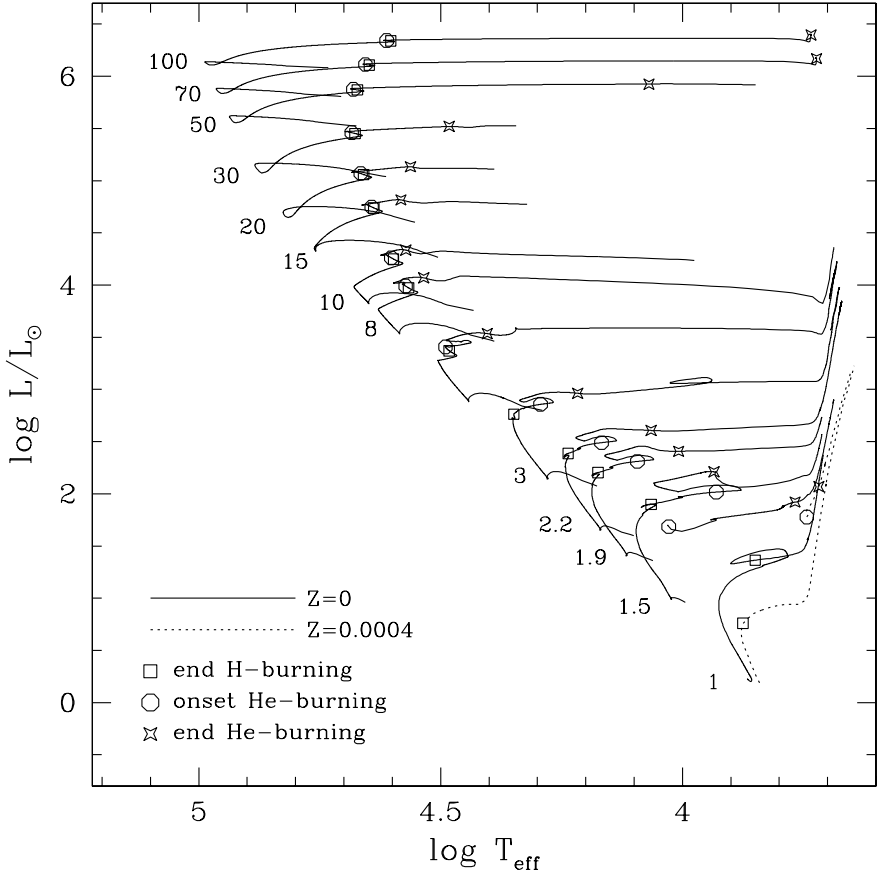}}
\end{minipage}
\hfill
\begin{minipage}{0.30\textwidth}
\caption{Zero-metal evolutionary tracks (solid lines) 
for selected initial masses (in $M_{\odot}$) as indicated.
The evolutionary track of the ($1\, M_{\odot}$, $Z=0.004$) model, 
calculated by Girardi et al. (2000),
is also shown for comparison (dotted line).}
\label{fig_hhe}
\end{minipage}
\end{figure*}

\subsection{The evolution on the H-R diagram}
{Let us now discuss the main evolutionary features 
of zero-metallicity stars
also in comparison to those of models with $Z \neq 0$.
The $Z=0$ stellar tracks 
in the H-R diagram are displayed in Fig.~\ref{fig_hhe},
where we mark the onset/end of major core burning phases.}

First, we can see that, as expected, the $Z=0$ tracks 
in the $\log L - \log T_{\rm eff}$ diagram are
systematically hotter and more luminous.
The latter feature {results in}  shorter nuclear lifetimes
(given in Table \ref{tab_time}) with respect to those of solar-composition
models.

Other {obvious} differences refer to 
(i) some morphological features of the tracks, and (ii) 
the relative excursion in  
in effective temperature  
characterising the nuclear H- and He-burning phases.

Concerning point (i), we notice that well-defined 
loops may show up on the  $Z=0$ tracks (e.g. 1 $M_{\odot}$ and 
3 $M_{\odot}$ models in Fig.~\ref{fig_hhe}). 
The physical interpretation of these
unusual features will be discussed in Sect.~\ref{sec_low}.

Concerning point (ii), it is worth making the following remarks.
As we can see, the locus of points 
corresponding to the termination of the main sequence (squares)
describes a path in the $\log T_{\rm eff}-\log L$ plane, which significantly  
bends towards lower effective temperatures at increasing stellar mass
(i.e. $M \ga  20 \,M_{\odot}$). 
{This} can be explained as due to the gradual
decrease of the mass exponent $a$ in the mass-luminosity relation
(i.e. $L \propto M^a$), as pointed out long ago by Stothers (1966).

Comparing the evolution of $1~M_{\odot}$ models with $Z=0$ and $Z=0.0004$, 
we see that once they experience powerful He-ignition at the tip
of the RGB, the subsequent quiescent core He-burning occurs  
either close to the Hayashi line ($Z=0.0004$), or 
{to an HB structure blueward of} the RR Lyrae instability strip 
(extending from $\log T_{\rm eff} \sim 3.75$ to 
$\log T_{\rm eff} \sim 3.85$; see Bono et al. 1995) {when} $Z=0$.

Moreover, in stars with $M \ga 5 M_{\odot}$, 
the onset of central He-burning (circles) 
occurs soon after hydrogen exhaustion 
in the core, and proceeds as these stars are evolving 
towards lower effective temperatures at nearly constant
luminosities. To this regard, a feature of particular interest 
is that the $Z=0$ models of intermediate mass 
($1.1\, M_{\odot} \la M_{\rm i} \la 7-8\, M_{\odot}$) 
start and complete their
core He-burning phase always in the bluest parts of the
tracks, i.e. quite far from their
Hayashi lines. This implies, for instance, that $Z=0$ intermediate-mass 
stars would never appear as either red-clump stars or blue-loop stars.

For even more massive models, 
the locus of points marking the end of the He-burning
phase (starred symbols in Fig.~\ref{fig_hhe}) also presents a systematic 
bending towards cooler regions at increasing stellar masses.
In particular, we find that massive models with 
$70 M_{\odot} \la M \la 100 M_{\odot}$ may reach their 
Hayashi lines already during the helium-burning phase,  
{ramaining there until} central carbon ignition.

\begin{figure*}
\begin{minipage}{0.69\textwidth}
\resizebox{\hsize}{!}{\includegraphics{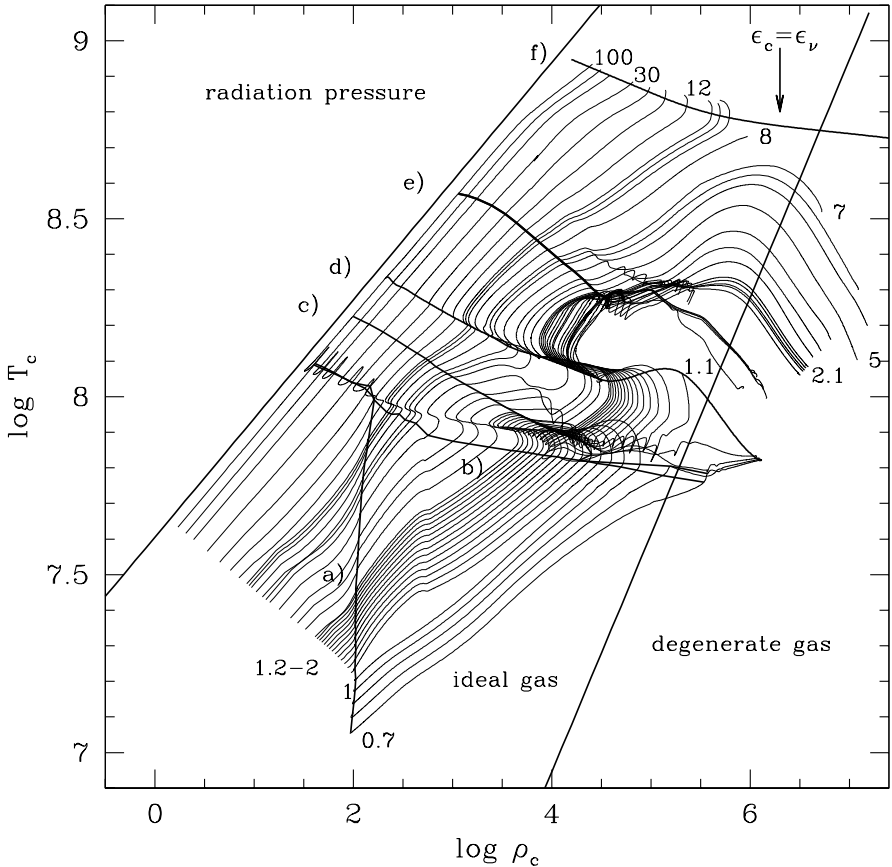}}
\end{minipage}
\hfill
\begin{minipage}{0.30\textwidth}
\caption{Evolution of central conditions (i.e. density 
and temperature) for the whole set 
of $Z=0$ models. Approximate relations are used to determine 
the boundaries (thick straight lines) 
of the regions in which  the equation of state is expected to be 
dominated by  different pressure components, i.e. degenerate gas, perfect
gas, and radiation pressure.
Labelled lines indicate point loci as a function
of the stellar mass corresponding to the onset of central H-burning (a);
onset of the 3-$\alpha$ reaction and hence the CNO-cycle (b); 
end of core H-burning (c);
onset of central He-burning (d); end of core He-burning (e); 
energy balance (f) between carbon burning 
and neutrino losses ($\epsilon_{\rm c}=\epsilon_{\nu}$).
A few values of the initial stellar masses are indicated (in $M_{\odot}$).}
\label{fig_tcroc}
\end{minipage}
\end{figure*}

In general, it should be remarked that whether or not a star spends some
part of its evolution on (or in proximity to) the Hayashi line may
have important implications. In fact, the corresponding development
of a deeply extended convective envelope may cause dredge-up events,
with  consequent changes in the surface chemical composition. 
This point will be discussed in Sect.~\ref{sec_chem}.  

\subsection{The evolution on the $T_{\rm c}-\rho_{\rm c}$ diagram}

In addition to the $\log L - \log T_{\rm eff}$ plane, 
another powerful diagnostic of the stellar structure is 
the central temperature -- central density diagram, as 
it illustrates the state of the gas in the innermost regions.
Figure \ref{fig_tcroc} displays the {evolutionary behaviour} 
in this plane by the models here considered.
Lines are drawn to mark the approximate boundaries of the regions in which
the equation of state is dominated by perfect gas, degenerate electron gas, 
and radiation pressure. 
Point loci as a function of the stellar 
mass are shown for relevant evolutionary stages.   

One can notice, in this figure, some typical behaviours
that depend mainly on the equation of state, and that are found in 
stellar models of any metallicity. For instance, 
in correspondence to the stages of gravitational contraction of 
stellar cores between nuclear burnings, 
the evolution of the central conditions in the domain of ideal gas
can be described by $T_{\rm c} \propto \rho_{\rm c}^{1/3}$,
as expected from homology considerations (e.g. Kippenhahn \& Weigert
1990).

However, the $T_{\rm c}-\rho_{\rm c}$ diagram of zero-metallicity stars
contain several unusual features worthy of {note}, such as:
the peculiar loci marking  the
onset of central H-burning (line a) and the onset of the 3-$\alpha$ reaction 
and the CNO-cycle (line b),
as well as some short-lived periods of core expansion and/or heating
(that are evident, for instance, soon above line b), and the unusually 
low mass limit ($1.1 \, M_\odot$) for stars to become degenerate before
He-ignition. Each one of these features will be discussed below.

\subsection{Up to the onset of the 3-$\alpha$ reaction}
\label{ssec_3alpha}
Owing to the complete lack of metals (and hence CNO elements)
during the ``initial'' stages of central H-burning, the 
only energy sources able to {satisfy} the condition 
of thermal equilibrium  are (i) gravitational contraction,
and (ii) nuclear burning via the p-p chains.
In low-mass stars, the p-p chain soon provides the nuclear energy 
necessary to slow down gravitational contraction, whereas at 
increasing mass this energy is reached only at
higher and higher temperatures.

The onset of the p-p reactions as a function of mass is
very inclined in the $T_{\rm c}-\rho_{\rm c}$ diagram (line a in
Fig.~\ref{fig_tcroc}).
Due to the {well known} rather weak dependence 
of the p-p reaction rates on 
temperature (i.e. $\epsilon_{\rm pp} \propto T^{4-6}$),  
the contraction of the central regions can easily proceed, 
so that high temperatures may be then attained during the H-burning
phase.

Once the central temperatures have increased up to typical values
of $\log T \sim 8$, the ignition of the 3-$\alpha$ reaction takes place. 
This occurrence marks a fundamental event, as it is represents the 
first significant production of metals in these stars.
As a consequence, the synthesis of primary
$^{12}$C leads to the activation of the CNO-cycle, which
then starts providing nuclear energy in competition with 
the p-p reactions (line b in Fig.~\ref{fig_tcroc}).

\begin{figure}
\resizebox{\hsize}{!}{\includegraphics{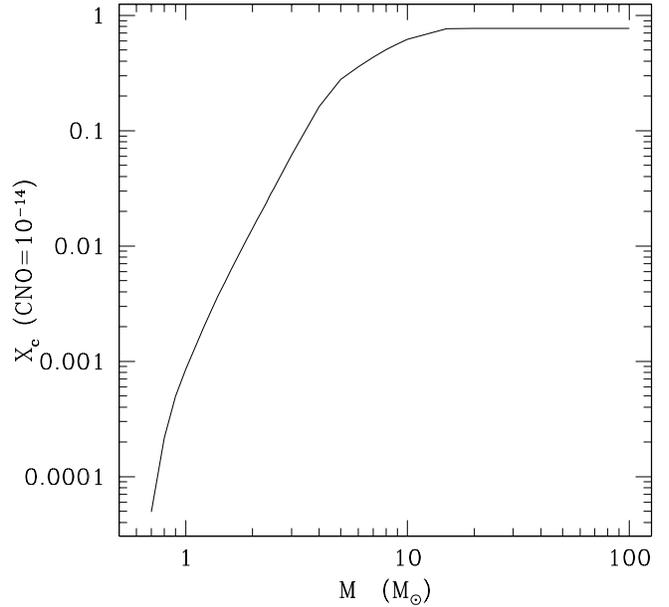}}
\caption{The central abundance of hydrogen when 
the fractional abundance of 
newly synthesised CNO elements grows up to $10^{-14}$.}
\label{fig_3a}
\end{figure}

The onset of the 3-$\alpha$ reaction occurs at earlier 
and earlier stages at increasing stellar mass (see Fig.~\ref{fig_3a}).
This occurs essentially because the 3-$\alpha$ reaction  
{requires much higher temperatures} than the p-p chain.
In the lowest mass models, the first 
production of $^{12}$C takes place towards the very end of the 
H-burning phase, and may not even occur for $M \la 0.8 \, M_{\odot}$.
In this latter case, all central hydrogen is entirely burnt via the
p-p chain. 
At a sufficiently high mass ($M \ge 20 M_\odot$ in our models),
the 3-$\alpha$ reaction ignites even {\em before}
the p-p reactions have slowed down the initial stellar contraction.
In these models then, H-burning simply proceeds {via} the CNO-cycle, 
{without} any significant phase of central burning via the p-p chain.  

The above features explain the striking change in the slope in the
curve corresponding to the onset of the H-burning phase as a 
function of the stellar mass (line a in Fig.~\ref{fig_tcroc}).
For models with $M \la 20 \,M_{\odot}$,  H-burning
starts above a nearly vertical line in the $T_{\rm c} - \rho_{\rm c}$ 
diagram (i.e. at varying $T_{\rm c}$ and for an almost constant 
$\rho_{\rm c}$), that is characteristic of ZAMS
stars that burn hydrogen predominantly through the p-p chain.
For models with $M \ga 20\, M_{\odot}$, H-burning starts
above a line of modest slope (i.e. a modest increase 
of $T_{\rm c}$ corresponds to a substantial decrease of $\rho_{\rm c}$),
that simply traces the minimum 
$T_{\rm c}$ and $\rho_{\rm c}$ necessary to ignite the 3-$\alpha$ 
reaction (and hence the CNO-cycle) in a core with $Y\sim0.23$.
This also explains why line (a) simply merges with line (b) at 
$M \ga 20 \,M_{\odot}$. 

\begin{figure}
\resizebox{\hsize}{!}{\includegraphics{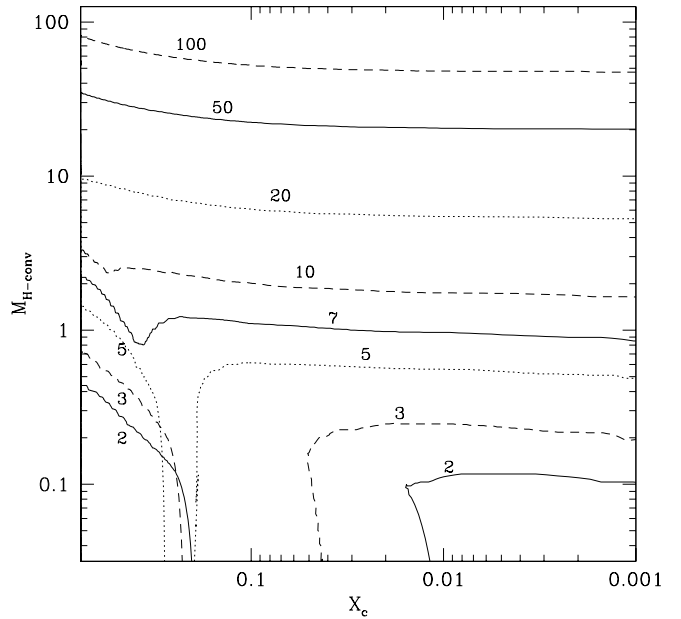}}
\caption{Evolution of core convection during central H-burning
for a few selected models with masses (in $M_{\odot}$) as indicated.}
\label{fig_convh}
\end{figure}

\subsection{Soon after the onset of the 3-$\alpha$ reaction}

Subsequent to the first carbon production, 
as the energy contribution of the CNO reactions increases 
(at the expense of the p-p reactions) the central regions are forced to 
expand, with consequent decrease of the central density (and 
central temperature in the most massive models). This
leads to the behaviour of the tracks in Fig.~\ref{fig_tcroc}, 
around the line (b).
 
The activation of the CNO-cycle
may result in the appearance of some peculiar features in the H-R diagram
of low-mass models (i.e. loops; see Sect.~\ref{sec_low}) and,
given its quite large temperature dependence
(i.e. $\epsilon_{\rm CNO} \propto T^{14-22}$), 
it also favours the development 
or a larger extension of core convection. 
For instance, in the case of 
the models with $M \sim 2 - 5 M_{\odot}$ displayed in
Fig.~\ref{fig_convh}, after the initial recession
of the convective core, central H-burning via the p-p
reactions occurs in radiative conditions.
Then, as the CNO cycle turns on,
central regions become convectively unstable again.
This effect is anticipated at increasing stellar mass, and eventually 
in more massive models ($M > 5 M_{\odot}$) the transition
{of the core from convective to radiative and back to convective} 
does not occur at all.

\begin{figure}
\resizebox{\hsize}{!}{\includegraphics{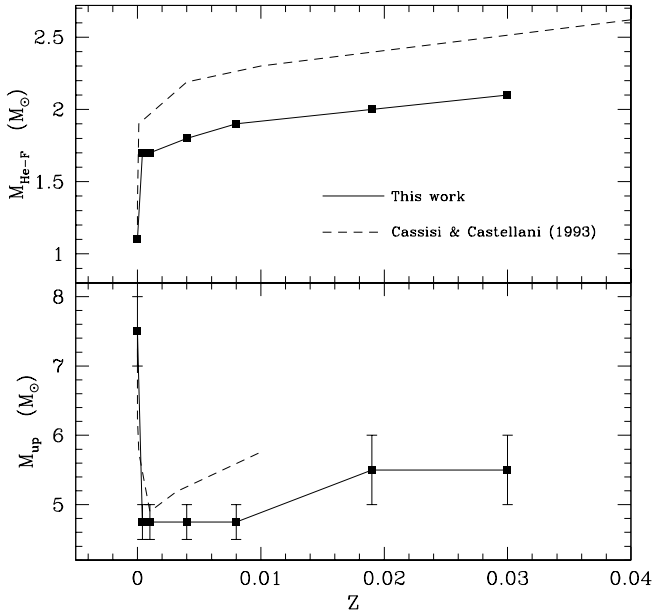}}
\caption{Critical masses $M_{\rm HeF}$ and $M_{\rm up}$ as a function
of the metallicity according to the predictions of Girardi
et al. (2000) for values of $Z \neq 0$, and those presently 
derived for $Z=0$. The results of Cassisi \& Castellani (1993; 
and references therein) are shown for comparison.}
\label{fig_mcrit}
\end{figure}

\subsection{The critical masses: $M_{\rm HeF}$ and $M_{\rm up}$}

From the location of the non-degenerate/degenerate boundary
and of the curves for He- and C-ignition  shown 
in Fig.~\ref{fig_tcroc}, 
it is possible to estimate the critical masses 
$M_{\rm HeF}$ and $M_{\rm up}$.
Figure \ref{fig_mcrit} illustrates the predicted trend of the
mass limits $M_{\rm HeF}$ and $M_{\rm up}$ as a function of the metallicity,
combining the results by Girardi et al. (2000) and those of the present 
work for $Z=0$.

Concerning $M_{\rm HeF}$, this is found to decrease with the metallicity,
dropping down to $\sim 1.1 M_{\odot}$ for $Z=0$.
This is the combined result of the larger convective cores during 
the MS phase at decreasing metallicities  
when the CNO cycle is the dominant energy source 
(up to $Z \sim 10^{-4}$, see CC93) 
and, at even lower $Z$, of the higher temperatures
reached in the centre when H-shell burning mainly occurs via the p-p 
reactions (as is the case for the $Z=0$ models).

Concerning  $M_{\rm up}$, its trend is not monotonic as it first
decreases with metallicity, reaches a minimum, and finally increases 
again up to $\ga 7\, M_{\odot}$ -- $8 \, M_{\odot}$ when $Z=0$.
This behaviour is essentially controlled by the mass of the He-core
left at the end of the MS. 
For CNO-dominated H-burning, lower metallicities
correspond to more concentrated energy sources and 
larger convective cores, which explains the initial
decrease of $M_{\rm up}$.
As soon as the energy contribution from the p-p reactions becomes
competitive with the CNO cycle, the central burning regions on the MS  
are more extended, while the sizes of the convective cores
tend to be smaller. This explains the presence of the minimum 
and the subsequent increase of $M_{\rm up}$ at decreasing $Z$.

Both behaviours of $M_{\rm HeF}$ and $M_{\rm up}$ are in agreement with 
the results of CC93. Their larger values of
$M_{\rm HeF}$ and $M_{\rm up}$ at given metallicity essentially reflect
the different treatment of convective boundaries (i.e. Schwarzschild 
criterion) with respect to ours (i.e. overshoot scheme; 
see Sect.~\ref{sec_inp}).

\begin{figure*}
\resizebox{\hsize}{!}{\includegraphics{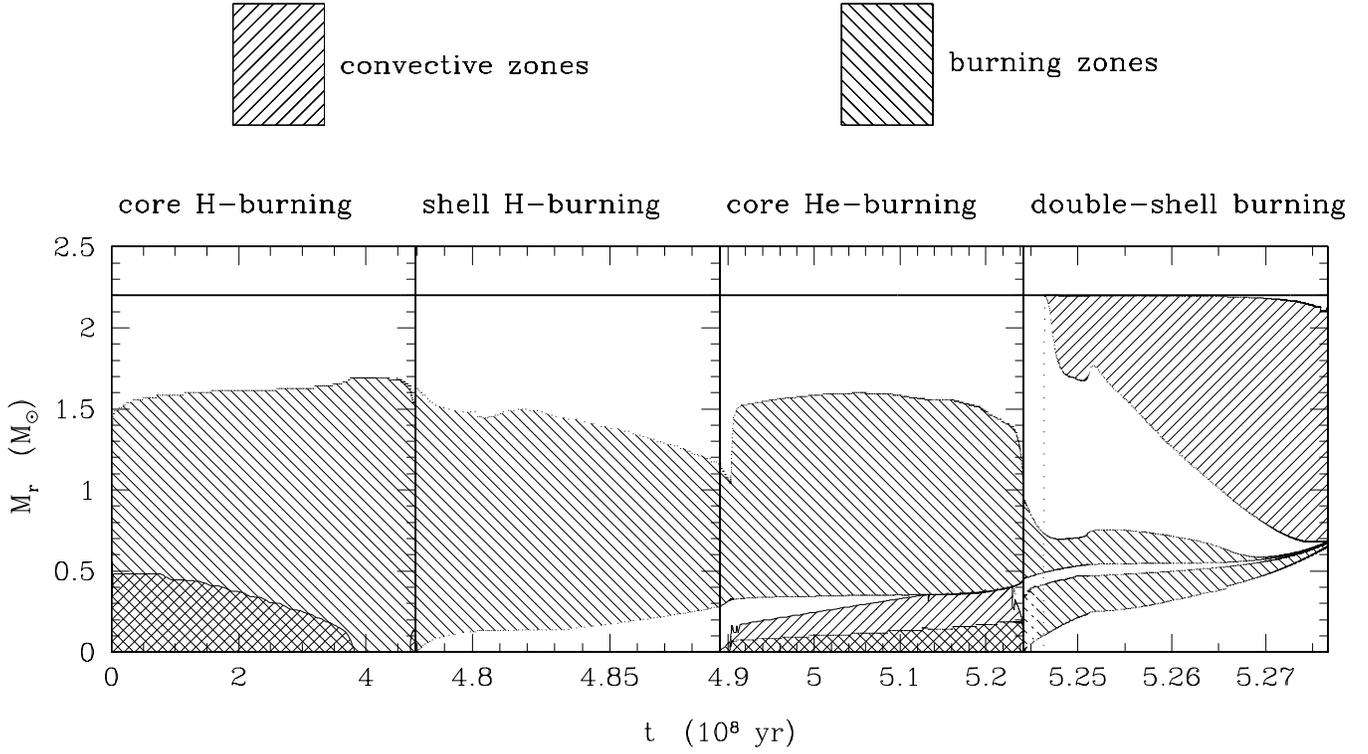}}
\caption{Kippenhahn diagrams showing the location (in fractional mass) 
of the burning and convective zones during the evolution of the
$2.2 \, M_{\odot}$, $Z=0$ model.}
\label{fig_conv}
\end{figure*}

\subsection{Growth of core convection during the He-burning phase}

Another remarkable evolutionary feature is
found in stars with masses $M \sim 1.2 - 2.5\, M_{\odot}$, and 
$M \ga 30 M_{\odot}$. 
During the central He-burning phase, the outer boundary of the
core overshooting zone is located very close to the bottom of 
the H-burning shell. 

{Initially, we found that,} if the convective
core was left to grow, it eventually reaches the H-shell and engulfs some
H-rich material, which is rapidly burnt in the core via the CNO cycle.
This causes a flash that expands the core, so that 
central He-burning weakens and the convective core recedes 
temporarily (in mass).
After the flash has occurred, the convective core 
starts growing again, possibly approaching the
bottom of the H-shell. This leads to 
quasi-periodic H-flashes, the related growth/recession of the core
resembling some sort of breathing convection.

However, this picture is likely to be physically unsound because of the 
the treatment adopted in the evolution code.
In fact, the H-flash is caused by the fact that the engulfed hydrogen
is first mixed throughout the convective core, and then burnt 
according to nuclear reaction rates which are mass-averaged all
over the convective layers. This procedure is  
a good approximation as long as the 
convective lifetimes are shorter than the nuclear lifetimes, which is 
usually the case for CNO-burning occurring in ``ordinary'' conditions.

\begin{figure*}
\begin{minipage}{0.69\textwidth}
\resizebox{\hsize}{!}{\includegraphics{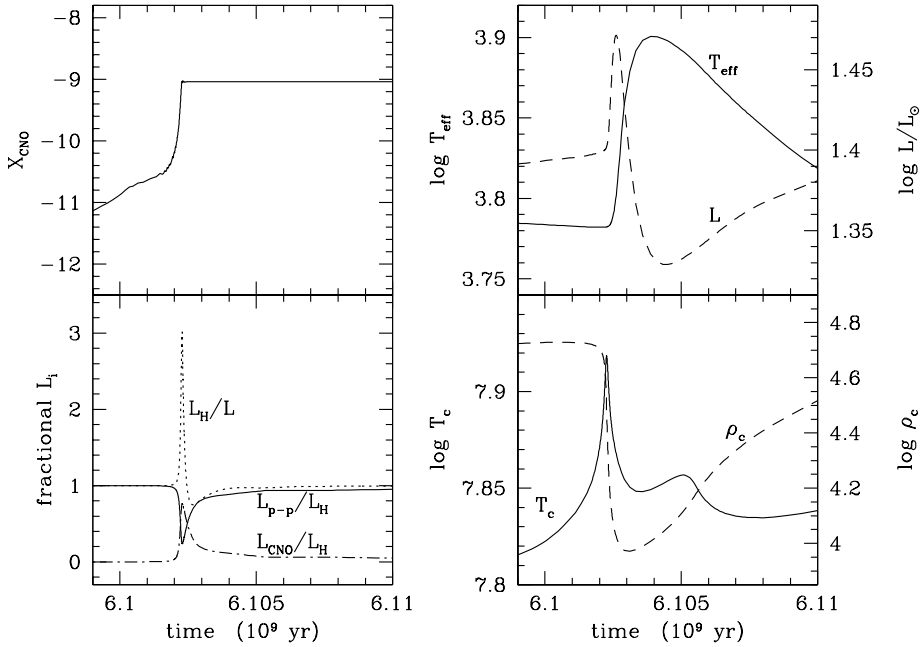}}
\end{minipage}
\hfill
\begin{minipage}{0.29\textwidth}
\caption{Physical properties of the 1 $M_{\odot}$ model
with $Z=0$ at the first onset of the 3-$\alpha$ reaction.
Top-left panel: central abundance of CNO elements.
Bottom-left panel: fractional energy contributions from
nuclear burnings, where $L$, $L_{\rm H}$, $L_{\rm CNO}$, 
and $L_{\rm pp}$ refer, respectively, to the surface stellar 
luminosity, the rate of energy production via H-burning, 
CNO cycle, and p-p chain.
Top-right panel: evolution of surface quantities as indicated.
Bottom-right panel: evolution of central density and temperature.}
\label{fig_loop1}
\end{minipage}
\end{figure*}

But at the very high temperatures 
($\sim 10^8$ K) reached at the bottom the H-burning shell,
the nuclear lifetimes involved in the CNO cycle (on the order of hours, days) 
can be comparable or even shorter than the typical 
convective timescales (on the order of months). 
Under these conditions, a correct approach would require a 
time-dependent solution scheme which couples simultaneously 
nucleosynthesis and mixing (see e.g. Schlattl et al. 2001)

Leaving the full analysis of this point to a future investigation,
we make the reasonable assumption in the present study 
that hydrogen burns locally at the  bottom
of the H-shell, before a complete mixing by core convection can occur.
Technically, this translates into the
condition that the maximum allowed extension of core convection
is set by the bottom of the H-burning shell.
  
Figure~\ref{fig_conv} illustrates the evolution of the convective and
burning regions of the 2.2~$M_\odot$ model. 
We can notice the approach of core convection
towards the inner location of the
H-burning shell, during the He-burning phase (third panel from left).
We can also notice the remarkable thickness (in mass) of the  
H-burning shell, which extends deeply inward during both shell 
H-burning and core He-burning phases.
This is due to the fact that hydrogen in the shell 
is mainly burnt via the p-p chain, which due to its low
temperature sensitivity, allows quite high temperatures to be 
attained at the bottom of the shell. 
At an age of $5.1\times10^8$~yr, our numerical treatment starts to limit the extension 
of the core convection. In this way, the evolution can 
be quietly followed up to the end of He-burning.

\section{Low- and intermediate-mass models}
\label{sec_low}
\label{sec_int}

\subsection{The first activation of the CNO cycle}
We have shown in Sect.~\ref{ssec_3alpha} that the onset of the
3-$\alpha$ reaction in low-mass models possibly occurs 
near the end of the central H-burning phase. 
This circumstance is marked by short-lived loops 
in the H-R diagram (see Figs.~\ref{fig_hr} and \ref{fig_hhe}).
We can better analyse this feature looking at Fig.~\ref{fig_loop1},
which displays how relevant properties of the 1 $M_{\odot}$ model vary 
during the brief time interval in which the loop shows up.

As soon as a CNO abundance of about $10^{-10,-9}$ is built at the centre
of the star, a sudden spike {arises} in the rate of energy generation
by nuclear H-burning. This occurs when the growing efficiency of the CNO
cycle exceeds that of the p-p reactions ($L_{\rm CNO}/L > L_{\rm pp}/L$),
which have dominated the energetics of the star up to then. 
The central conditions react to this sudden increase of energy production:
the innermost regions expand so that both $T_{\rm c}$ and $\rho_{\rm c}$  
decrease.

%
\subsection{The evolution on the RGB}
\label{ssec_rgb}
As already pointed out by D'Antona (1982) and CC93,
a prominent feature of zero-metallicity low-mass models 
is that, for given stellar
mass, the tip of the RGB is reached with
a larger core mass and much fainter luminosity with
respect to models with non-zero metallicity 
(see, for instance,  the $1\, M_{\odot}$ tracks for $Z=0$ and $Z=0.0004$ 
in Fig.~\ref{fig_hhe}).
 
The peculiar behaviour
of the luminosity as a function of the core mass 
($M_{\rm c} - L$ relation) displayed 
by RGB models with $Z=0$ can be explained as the combined effect 
of the lower mean molecular weight ($\mu$), and the dominant 
nuclear energy source (i.e. p-p chain).
{As is shown by Marigo (2000)}, the $M_{\rm c} - L$ relation 
at $Z=0$ can be remarkably well reproduced by the homology prediction
$L \propto (\mu M_{\rm c})^{\delta}$ with $\delta \sim 4.5$, 
the value of the exponent being essentially determined by the
typical temperature dependence of the p-p reactions.
For dominating CNO cycle, we would get  
$\delta \sim 7-8$, which well accounts 
for the luminosity evolution of RGB models with $Z > 0$.

\begin{figure}
\resizebox{\hsize}{!}{\includegraphics{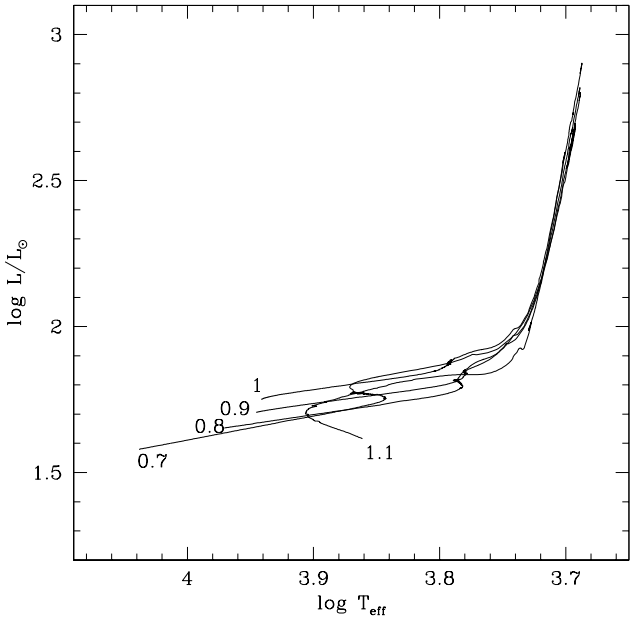}}
\caption{H-R tracks of $Z=0$ low-mass models  extending from
the ZAHB up to the beginning of the TP-AGB phase.}
\label{fig_hb}
\end{figure}


\subsection{The evolution on the HB}
\label{ssec_hb}
In principle, as a low-mass star 
settles on the ZAHB, it starts to quiescently burn 
helium in the core with an abundance somewhat 
lower than the original one, since a small amount 
of helium, $\Delta Y_{\rm c}^{\rm HeF}$,
has been already converted into carbon during 
the flash at the tip of the RGB.

\begin{table}
\caption{Characteristics of low-mass stars at the He-flash.}
\label{tab_heflash}
\begin{tabular}{llll}
\noalign{\smallskip}\hline\noalign{\smallskip}
$M$ & $M_{\rm c}^{\rm HeF}$ &
$\Delta E_{\rm RGBt}^{\rm ZAHB}$ & $\Delta Y_{\rm c}^{\rm HeF}$ \\
$(\Msun)$ & $(\Msun)$ & $(10^{49}\,{\rm erg})$ & $$ \\
\noalign{\smallskip}\hline\noalign{\smallskip}
0.7 & 0.5015 & 2.573 & $-$0.0325 \\
0.8 & 0.4991 & 2.549 & $-$0.0322 \\
0.9 & 0.4951 & 2.497 & $-$0.0315 \\
1.0 & 0.4882 & 2.400 & $-$0.0303 \\
1.1 & 0.3683 & 0.989 & $-$0.0100 \\
\noalign{\smallskip}\hline\noalign{\smallskip}
\end{tabular}
\end{table}

We have computed this 
quantity for each evolutionary track, by simply assuming
that the flash provides just the energy
necessary to lift the electron degeneracy from the core. 
In practice, this energy  
can be evaluated as the difference in total energy between the last 
RGB-tip configuration and the initial ZAHB one, 
$\Delta E_{\rm RGBt}^{\rm ZAHB}$. Then, 
the mass of burnt helium, and hence $\Delta Y_{\rm c}^{\rm HeF}$, 
follows straightforwardly from the $Q$-value of 3-$\alpha$ reactions.
Table~\ref{tab_heflash} presents some relevant 
characteristics of the models at the stages of He-flash and ZAHB.
We can notice that the predicted fractional abundance of helium
is burnt in the flash amounts to few percent (typically $3\, \%$). 

The subsequent evolution on the HB (see Fig.~\ref{fig_hb}) 
is quite similar to that of
Pop-II models. However, as already pointed out by Cassisi et al. (1996),
all low-mass $Z=0$ stars burn their central He at 
quite high effective temperatures (with $\log T_{\rm eff} \ga 3.90$), 
so that a negligible fraction of RR Lyrae pulsators {on the ZAHB} 
is expected.
This aspect is further commented on Sect.~\ref{sec_strip}.

\begin{figure*}
\begin{minipage}{0.69\textwidth}
\resizebox{\hsize}{!}{\includegraphics{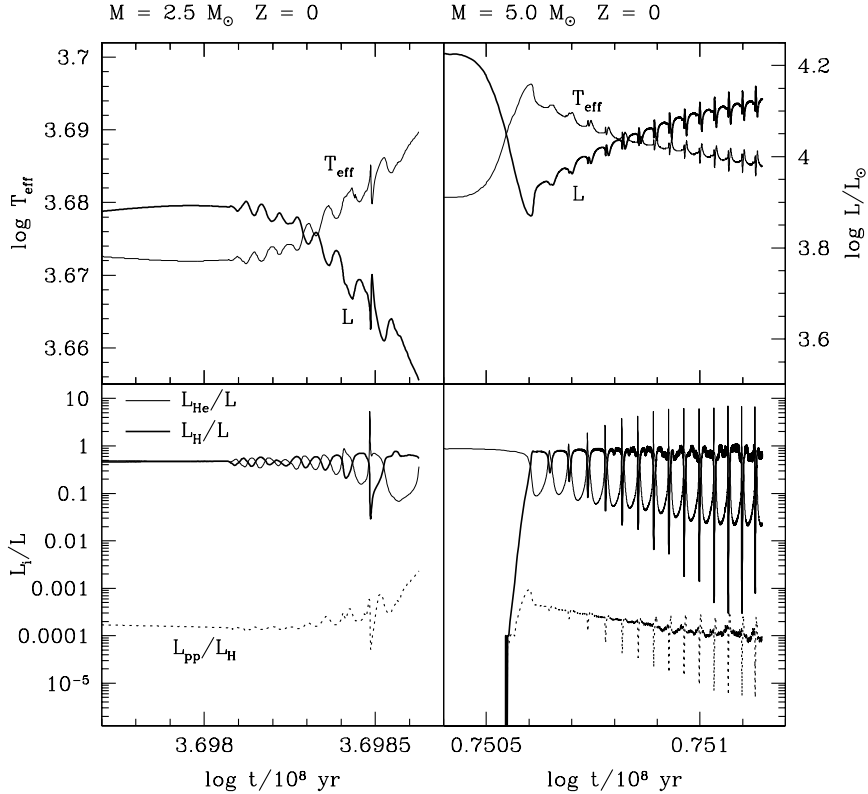}}
\end{minipage}
\hfill
\begin{minipage}{0.29\textwidth}
\caption{Evolutionary properties during the first thermal instabilities of the 
He-burning shell, for two $Z=0$ AGB models with initial masses of
$2.5 \, M_{\odot}$, and $5.0 \, M_{\odot}$. See text for further explanation.}
\label{fig_agb}
\end{minipage}
\end{figure*}

\subsection{The AGB phase and the occurrence of thermal pulses}
\label{ssec_agb}

The evolutionary properties of metal deficient AGB stars have been 
analysed in the past by Fujimoto et al. (1984), Chieffi \& Tornamb\'e
(1984), Cassisi et al. (1996),  
and more recently by Dom\'\i nguez et al. (1999).
A crucial question addressed in these works 
is about the possible occurrence  of the He-shell flashes (thermal pulses),
which are usually found in AGB models of various (but not zero) metallicities.

According to the semi-analytical investigation of  Fujimoto et al. (1984)
on the stability of nuclear burning shells,  
thermal pulses are {conditional on the} existence of thresholds in
core mass $M_{\rm c}$, and CNO abundance $Z_{\rm CNO}$ in the envelope.
To summarise, He-shell flashes are expected to occur when
i) $M_{\rm c} < 0.73\, M_{\odot}$ for any value of $Z_{\rm CNO}$,
and ii)  $M_{\rm c} > 0.73\, M_{\odot}$ if 
$Z_{\rm CNO} > Z_{\rm CNO}^{\rm crit}$, the critical lower limit
depending on $M_{\rm c}$.
The dependence on $Z_{\rm CNO}$ can be qualitatively understood considering
that in the absence (or strong deficit) of CNO elements in the envelope,
the H-burning shell {itself} 
produces new carbon via the 3-$\alpha$ reaction,
so that the CNO cycle can operate. It follows that the H-
and He-burning shells are not energetically decoupled, and proceed outward 
in mass at the same rate (see also Chieffi \& Tornamb\'e 1984).
Since He-rich material is not accumulated in the inert intershell buffer,
but rather steadily burnt by the underlying He-shell, 
the occurrence of thermal pulses is thus prevented.

In this study we present the results for two intermediate-mass models,
with initial zero metallicity and  
masses of $2.5 \, M_{\odot}$ and $5 \, M_{\odot}$,
which are made evolve through the initial stages of the AGB phase,
including a few episodes of thermal instabilities  of the He-burning
shell. Whether such events may be ascribed to true He-shell flashes
is discussed {below}.

As the $2.5 \, M_{\odot}$ and 
$5 \, M_{\odot}$ models reach their Hayashi tracks (E-AGB phase)
after central helium exhaustion, envelope convection progressively 
moves inward {in both cases, and extends} into regions which have 
experienced earlier nucleosynthesis.
The so-called ``second dredge-up'' takes place, but with a substantial
difference between the two models. This can be appreciated by looking at
Table~\ref{tab_du}, which presents the predicted surface chemical 
compositions after the second dredge-up event for all our models.

\begin{center}
\begin{table*}[landscape]
\caption{Surface chemical compositions (in mass fraction) after the 
1$^{\rm st}$ and 2$^{\rm nd}$ dredge-up events.}
\label{tab_du}
\rotatebox{90}{
\begin{tabular}{rccccccccccccc}
\noalign{\smallskip}\hline\noalign{\smallskip}
$M/M_{\odot}$ &      H  &    $^3$He  &      $^4$He  &  $^{12}$C   &    $^{13}$C   & 
   $^{14}$N   &    $^{15}$N  &     $^{16}$O   &    $^{17}$O  &     $^{18}$O  & $^{22}$Ne  & $^{24+25}$Mg & $^{20}$Ne \\
\noalign{\smallskip}\hline\noalign{\smallskip}
\multicolumn{11}{l}{Initial:}   \\
  all &  0.770 & 2.46 $10^{-5}$ &  0.230 & 0 & 0 & 0 & 0 & 0 & 0 & 0 & 0 & 0 & 0
 \\
\noalign{\smallskip}\hline\noalign{\smallskip}
\multicolumn{11}{l}{After the first dredge-up:} \\
 0.7 &  0.769 & 8.63$\,10^{-5}$ &  0.231 & 0 & 0 & 0 & 0 & 0 & 0 & 0 & 0 & 0 & 0 \\
 0.8 &  0.768 & 1.40$\,10^{-4}$ &  0.232 & 0 & 0 & 0 & 0 & 0 & 0 & 0 & 0 & 0 & 0 \\
 0.9 &  0.768 & 1.54$\,10^{-4}$ &  0.232 & 0 & 0 & 0 & 0 & 0 & 0 & 0 & 0 & 0 & 0 \\
 1.0 &  0.768 & 1.17$\,10^{-4}$ &  0.232 & 0 & 0 & 0 & 0 & 0 & 0 & 0 & 0 & 0 & 0 \\
 1.1 &  0.769 & 8.29$\,10^{-5}$ &  0.231 & 0 & 0 & 0 & 0 & 0 & 0 & 0 & 0 & 0 & 0 \\
\noalign{\smallskip}\hline\noalign{\smallskip}
\multicolumn{11}{l}{After the second dredge-up:} \\
 1.2 &  0.769 & 6.49$\,10^{-4}$ &  0.231 & 0 & 0 & 0 & 0 & 0 & 0 & 0 & 0 & 0 & 0 \\
 1.3 &  0.767 & 4.89$\,10^{-4}$ &  0.233 & 0 & 0 & 0 & 0 & 0 & 0 & 0 & 0 & 0 & 0 \\
 1.4 &  0.769 & 4.72$\,10^{-4}$ &  0.231 & 0 & 0 & 0 & 0 & 0 & 0 & 0 & 0 & 0 & 0 \\
 1.5 &  0.767 & 3.45$\,10^{-4}$ &  0.233 & 0 & 0 & 0 & 0 & 0 & 0 & 0 & 0 & 0 & 0 \\
 1.6 &  0.767 & 3.16$\,10^{-4}$ &  0.232 & 0 & 0 & 0 & 0 & 0 & 0 & 0 & 0 & 0 & 0 \\
 1.7 &  0.767 & 2.75$\,10^{-4}$ &  0.233 & 0 & 0 & 0 & 0 & 0 & 0 & 0 & 0 & 0 & 0 \\
 1.8 &  0.769 & 2.64$\,10^{-4}$ &  0.231 & 0 & 0 & 0 & 0 & 0 & 0 & 0 & 0 & 0 & 0 \\
 1.9 &  0.768 & 2.19$\,10^{-4}$ &  0.232 & 0 & 0 & 0 & 0 & 0 & 0 & 0 & 0 & 0 & 0 \\
 2.0 &  0.768 & 2.02$\,10^{-4}$ &  0.232 & 0 & 0 & 0 & 0 & 0 & 0 & 0 & 0 & 0 & 0 \\
 2.1 &  0.704 & 1.07$\,10^{-4}$ &  0.296 & 0 & 0 & 0 & 0 & 0 & 0 & 0 & 0 & 0 & 0 \\
 2.2 &  0.697 & 9.59$\,10^{-5}$ &  0.303 & 0 & 0 & 0 & 0 & 0 & 0 & 0 & 0 & 0 & 0 \\
 2.3 &  0.691 & 8.13$\,10^{-5}$ &  0.309 & 0 & 0 & 0 & 0 & 0 & 0 & 0 & 0 & 0 & 0 \\
 2.4 &  0.690 & 6.82$\,10^{-5}$ &  0.310 & 0 & 0 & 0 & 0 & 0 & 0 & 0 & 0 & 0 & 0 \\
 2.5 &  0.681 & 5.97$\,10^{-5}$ &  0.319 & 0 & 0 & 0 & 0 & 0 & 0 & 0 & 0 & 0 & 0 \\
 2.7 &  0.669 & 5.04$\,10^{-5}$ &  0.331 & 4.17$\,10^{-21}$ & 1.34$\,10^{-21}$ & 4.95$\,10^{-19}$ & 2.02$\,10^{-23}$ & 1.28$\,10^{-20}$ & 1.33$\,10^{-22}$ & 7.16$\,10^{-26}$ & 2.43$\,10^{-35}$ & 1.62$\,10^{-50}$ & 7.70$\,10^{-38}$ \\
 3.0 &  0.660 & 4.10$\,10^{-5}$ &  0.340 & 1.13$\,10^{-18}$ & 3.62$\,10^{-19}$ & 1.28$\,10^{-16}$ & 5.19$\,10^{-21}$ & 3.15$\,10^{-18}$ & 2.82$\,10^{-20}$ & 1.60$\,10^{-23}$ & 4.25$\,10^{-32}$ & 6.91$\,10^{-46}$ & 3.19$\,10^{-34}$  \\
 3.5 &  0.649 & 2.99$\,10^{-5}$ &  0.351 & 8.06$\,10^{-16}$ & 2.59$\,10^{-16}$ & 8.58$\,10^{-14}$ & 3.46$\,10^{-18}$ & 1.97$\,10^{-15}$ & 1.41$\,10^{-17}$ & 8.69$\,10^{-21}$ & 2.06$\,10^{-28}$ & 1.86$\,10^{-40}$ & 6.19$\,10^{-30}$ \\
 4.0 &  0.644 & 2.29$\,10^{-5}$ &  0.356 & 6.33$\,10^{-14}$ & 2.03$\,10^{-14}$ & 6.17$\,10^{-12}$ & 2.46$\,10^{-16}$ & 1.31$\,10^{-13}$ & 7.41$\,10^{-16}$ & 5.23$\,10^{-19}$ & 5.33$\,10^{-26}$ & 7.42$\,10^{-37}$ & 4.13$\,10^{-27}$ \\
 5.0 &  0.637 & 1.56$\,10^{-5}$ &  0.363 & 3.58$\,10^{-9}$ & 7.81$\,10^{-12}$ & 6.06$\,10^{-10}$ & 1.57$\,10^{-14}$ & 4.54$\,10^{-12}$ & 3.90$\,10^{-14}$ & 5.11$\,10^{-15}$ & 5.22$\,10^{-19}$ & 6.68$\,10^{-28}$ & 1.14$\,10^{-23}$ \\
 6.0 &  0.632 & 1.11$\,10^{-5}$ &  0.368 & 2.63$\,10^{-7}$ & 1.19$\,10^{-11}$ & 9.35$\,10^{-10}$ & 2.41$\,10^{-14}$ & 7.36$\,10^{-11}$ & 5.91$\,10^{-14}$ & 4.67$\,10^{-13}$ & 2.96$\,10^{-16}$ & 1.66$\,10^{-23}$ & 4.73$\,10^{-21}$ \\
 6.5 &  0.626 & 9.25$\,10^{-6}$ &  0.374 & 7.07$\,10^{-7}$ & 1.46$\,10^{-11}$ & 1.13$\,10^{-9}$ & 2.90$\,10^{-14}$ & 2.99$\,10^{-10}$ & 7.37$\,10^{-14}$ & 2.21$\,10^{-12}$ & 1.95$\,10^{-15}$ & 2.63$\,10^{-22}$ & 4.09$\,10^{-20}$ \\
 7.0 &  0.668 & 8.74$\,10^{-6}$ &  0.332 & 1.26$\,10^{-13}$ & 4.04$\,10^{-14}$ & 1.09$\,10^{-11}$ & 4.30$\,10^{-16}$ & 2.08$\,10^{-13}$ & 8.57$\,10^{-16}$ & 7.51$\,10^{-19}$ & 6.02$\,10^{-26}$ & 2.75$\,10^{-36}$ & 9.03$\,10^{-27}$ \\
 8.0 &  0.681 & 7.49$\,10^{-6}$ &  0.319 & 9.66$\,10^{-14}$ & 5.68$\,10^{-17}$ & 8.08$\,10^{-12}$ & 6.32$\,10^{-19}$ & 1.47$\,10^{-13}$ & 1.41$\,10^{-18}$ & 1.15$\,10^{-21}$ & 9.09$\,10^{-30}$ & 6.05$\,10^{-42}$ & 4.43$\,10^{-31}$ \\
 8.3 &  0.687 & 7.00$\,10^{-6}$ &  0.313 & 1.19$\,10^{-14}$ & 3.81$\,10^{-15}$ & 1.01$\,10^{-12}$ & 3.98$\,10^{-17}$ & 1.88$\,10^{-14}$ & 7.39$\,10^{-17}$ & 6.71$\,10^{-20}$ & 1.95$\,10^{-27}$ & 2.12$\,10^{-38}$ & 2.36$\,10^{-28}$ \\
 9.0 &  0.770 & 4.30$\,10^{-5}$ &  0.230 & 0 & 0 & 0 & 0 & 0 & 0 & 0 & 0 & 0 & 0 \\
 9.5 &  0.770 & 4.30$\,10^{-5}$ &  0.230 & 0 & 0 & 0 & 0 & 0 & 0 & 0 & 0 & 0 & 0 \\
 10.0 &  0.770 & 4.25$\,10^{-5}$ &  0.230 & 0 & 0 & 0 & 0 & 0 & 0 & 0 & 0 & 0 & 0 \\
 12.0 &  0.770 & 3.63$\,10^{-5}$ &  0.230 & 0 & 0 & 0 & 0 & 0 & 0 & 0 & 0 & 0 & 0 \\
 15.0 &  0.770 & 2.47$\,10^{-5}$ &  0.230 & 0 & 0 & 0 & 0 & 0 & 0 & 0 & 0 & 0 & 0 \\
 20.0 &  0.770 & 1.66$\,10^{-5}$ &  0.230 & 0 & 0 & 0 & 0 & 0 & 0 & 0 & 0 & 0 & 0  \\
 30.0 &  0.770 & 1.02$\,10^{-5}$ &  0.230 & 0 & 0 & 0 & 0 & 0 & 0 & 0 & 0 & 0 & 0 \\
 50.0 &  0.770 & 5.95$\,10^{-6}$ &  0.230 & 0 & 0 & 0 & 0 & 0 & 0 & 0 & 0 & 0 & 0 \\
 70.0 &  0.713 & 1.06$\,10^{-6}$ &  0.287 & 7.65$\,10^{-12}$ &  2.45$\,10^{-12}$ &   6.39$\,10^{-10}$ &   2.53$\,10^{-14}$ & 1.30$\,10^{-11}$ &
5.29$\,10^{-14}$ & 5.34$\,10^{-17}$ &  5.11$\,10^{-24}$ & 2.08$\,10^{-14}$ &
2.58$\,10^{-14}$ \\
 100.0 &  0.640 & 6.39$\,10^{-7}$ &  0.360 & 2.27 $\,10^{-11}$ & 7.25$\,10^{-12}$ &  1.87$\,10^{-9}$ & 7.21$\,10^{-14}$ & 3.53 $\,10^{-11}$ & 1.33$\,10^{-13}$ &  1.33$\,10^{-16}$ &  2.03$\,10^{-23}$ &  1.42$\,10^{-32}$  &  1.13$\,10^{-23}$ \\
\noalign{\smallskip}\hline\noalign{\smallskip}
\end{tabular}
}
\end{table*}
\end{center}

In the $2.5 \, M_{\odot}$ model the envelope penetrates into the 
chemical profile left by the recession of the convective core during 
the MS, and into regions previously affected by the p-p chain, 
so that the surface enrichment is essentially in helium
with no trace of heavier metals
(see also Fig.~\ref{fig_conv} for a similar case).
In the $5 \, M_{\odot}$ model the envelope penetration 
proceeds further, reaching  down
to the former location of the H-burning shell 
(which temporarily extinguishes), so that
newly synthesized CNO elements are also
dredged-up to the surface.
After the occurrence of the second dredge-up, the evolution of  
these AGB models proceeds quite differently.

As we can see from Fig.~\ref{fig_agb}, the $2.5 \, M_{\odot}$ {model}
experiences rather
weak fluctuations in the surface luminosity, reflecting the alternation
between the H- and He-burning shells in providing nuclear energy 
to the star. The long-term behaviour of the surface properties indicates
the model is evolving down on its Hayashi track, i.e. 
to decreasing luminosity and increasing effective temperature.
As already mentioned, the CNO surface abundance is zero, so that
the H-burning shell must {itself} produce the necessary $^{12}$C to
sustain the CNO-cycle.  The latter is actually the dominant energy
source during most part of the AGB evolution followed 
by our calculations (bottom-left panel of Fig.~\ref{fig_agb}).

The $5 \, M_{\odot}$ model suffers well-defined He-shell flashes,
which drive the features usually {found} in the light curves 
of AGB stars.
Namely, over a pulse cycle, the luminosity {undergoes} the
initial post-flash rapid dip and peak, which is followed by the 
low-luminosity dip, and the final recovering 
of the luminosity level that characterizes the quiescent shell 
H-burning (see, for instance, Boothroyd \& Sackmann 1988).
In this case, the long-term  evolution is characterised by the increase 
of the surface luminosity and slight decrease of the effective 
temperature, i.e. the model is climbing on the AGB.
The envelope CNO abundance (by mass) left after the completion of the second
dredge-up is $Z_{\rm CNO} = 4.198 \times 10^{-9}$. 

As for the 2.5 $M_{\odot}$ model, the CNO cycle provides
most of the stellar energy during the stages of quiescent H-burning 
(bottom-right panel of Fig.~\ref{fig_agb}).
However, it should be noticed that a fraction of the CNO nuclei involved
as catalysts in the CNO cycle are not brought up to
the surface by the second dredge-up, but are synthesized 
in situ by the H-burning shell {via the $3-\alpha$ process}.
This can be seen by comparing the total CNO abundance (by number) 
in the envelope, $Z_{\rm CNO}$,
and that, for instance, at the point of maximum nuclear efficiency 
within the H-burning shell, $Z_{\rm CNO}^{\rm shell}$. 
If there were not production of $^{12}$C {inside} the shell,
we should have  $Z_{\rm CNO} = Z_{\rm CNO}^{\rm shell}$, whereas
we find $Z_{\rm CNO} < Z_{\rm CNO}^{\rm shell}$, with 
$Z_{\rm CNO}^{\rm shell}$ ranging from 
$\sim 2\times 10^{-8}$ to  $\sim 6.5 \times 10^{-8}$ at 
subsequent stages immediately preceding a thermal pulse.

Let us now check whether our results can be recovered  
within the  Fujimoto scheme.
The $2.5 \, M_{\odot}$  and $5 \, M_{\odot}$ models have core masses
of about $0.69 \, M_{\odot}$ and $0.87 \, M_{\odot}$, and $Z_{\rm CNO}$
in the envelope equal to 0 and $\sim 4.198 \times 10^{-9}$, respectively.
According to Fujimoto et al. (1984) 
the lowest mass model is expected to suffer thermal pulses, 
since $M_{\rm c} < M_{\rm c}^{\rm crit}$ (regardless of $Z_{\rm CNO}$), 
whereas the most massive one should not, since it is located
inside the region of the plane $M_{\rm c}-Z_{\rm CNO}$ 
where thermal pulses are prohibited (see their figure~5).
Thus, our results would seem to disagree with the Fujimoto scheme.
This is perhaps not surprising as the predictions by Fujimoto 
et al. (1984) {are derived} from a semi-analytical method, in which the 
input physics -- adopted for the integrations of deep envelope 
structures -- may differ from those presently used in our stellar code,  
as well as for other model prescriptions, 
{such as} the initial helium abundance ($Y=0.25$ in Fujimoto et al.,
$Y=0.23$ in our models). Other differences may also arise from
the use of analytical approximations in the Fujimoto procedure. 
  
\begin{figure*}
\resizebox{\hsize}{!}{\includegraphics{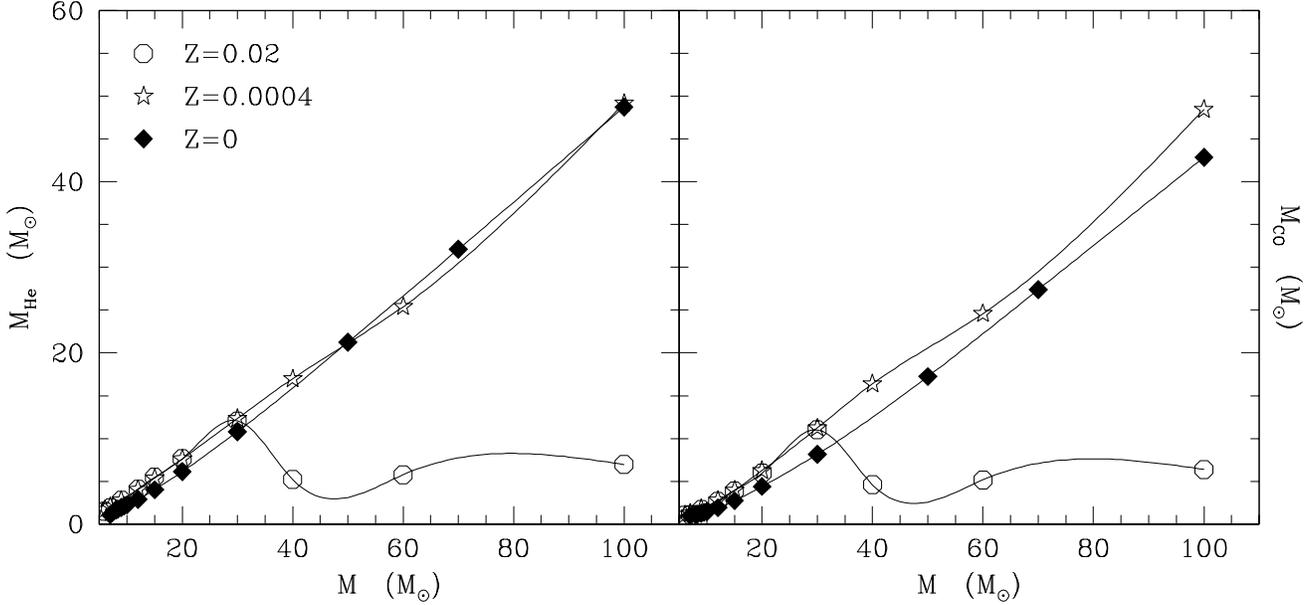}}
\caption{Masses of the He core 
($M_{\rm He}$) and C-O core ($M_{\rm CO}$) at the central carbon
ignition in zero-metallicity massive models, compared with predictions
for $Z=0.02$, and $Z=10^{-4}$.}
\label{fig_mheco}
\end{figure*}

However, 
even if our results do not strictly obey the quantitative thresholds
in $M_{\rm c}$ and $Z_{\rm CNO}$ as pointed out by Fujimoto et al. (1984),
a qualitative agreement in the behaviour of AGB models with initial 
zero metallicity can be found.
In particular, we confirm the prediction that the occurrence of thermal
pulses in massive AGB models is {conditional on} the pre-existence of
some CNO abundance in the envelope, that in the case of our
$5 M_{\odot}$ model is supplied by the second dredge-up during the E-AGB.
This finding is also supported by the evolutionary calculations
of zero-metallicity intermediate-mass stars 
by Chieffi \& Tornamb\'e (1984) and Dom\'\i nguez et al. (1999).
 
The strength of thermal instabilities should also depend on 
the $Z_{\rm CNO}$ value in the envelope, as this quantity controls the 
degree of thermal coupling between the H- and He-burning shell.
In models with initial zero metallicity, we expect that
the more $Z_{\rm CNO}$ is dredged-up, 
the less $^{12}$C is synthesised by the 3-$\alpha$ 
reaction in H-shell during a interpulse period,  
hence the stronger the He-shell flash should be.

\begin{table}
\caption{Core masses and central C/O ratio at the stage of C-ignition.}
\label{tab_mcore}
\begin{tabular}{llll}
\noalign{\smallskip}\hline\noalign{\smallskip}
$M/M_{\odot}$ & $M_{\rm He}/M_{\odot}$   &   $M_{\rm CO}/M_{\odot}$ &  
$(X_{\rm C}/X_{\rm O})_{\rm c}$ \\
\noalign{\smallskip}\hline\noalign{\smallskip}
7.0 & 1.09 & 0.98 & 0.715 \\
8.0 & 1.66 & 1.13 & 0.688 \\
8.3 & 1.75 & 1.18 & 0.677 \\
9.0 & 1.96 & 1.29 & 0.658 \\
9.50 & 2.11 & 1.38 & 0.643 \\
10.0 & 2.25 & 1.47 & 0.630 \\
12.0 & 2.90 & 1.92 & 0.574 \\
15.0 & 4.01 & 2.73 & 0.506 \\
20.0 & 6.13 & 4.40 & 0.422 \\
30.0 & 10.79 & 8.17 & 0.330 \\
50.0 & 21.25 & 17.23 & 0.234 \\
70.0 & 32.10 & 27.39 & 0.187 \\
100.0 & 48.73 & 42.84 & 0.149 \\
\noalign{\smallskip}\hline\noalign{\smallskip}
\end{tabular}
\end{table}

\section{Massive models}
\label{sec_mas}
%
\subsection{Core masses}
In Figure \ref{fig_mheco} and Table \ref{tab_mcore} we report the masses
of the He and C-O cores of the zero-metallicity models with
$M > M_{\rm up} \sim 7 \, M_{\odot}$, which are expected to
end their evolution as supernovae.

These quantities represent  fundamental information 
for both hydrodynamic calculations of the explosion event (see, for instance, 
Umeda et al. 2000), and  synthetic derivation of the
supernova yields (see the procedure described by Portinari et al. 1998). 

Comparing the results for $Z=0$ with those for other metallicities,
we can notice that: i) for initial masses $\la 25-30 \, M_{\odot}$ 
the core masses are almost independent of metallicity; ii) the major 
differences show up at higher masses, being largely determined by the
effect of mass loss that is expected to be less
efficient at lower $Z$ (and not applied to the $Z=0$ models);
iii) in the case of the $Z=0$ models 
evolved at constant mass, both $M_{\rm He}$ and $M_{\rm CO}$
follow a {nearly} linear relation with the stellar mass.   

\subsection{The Eddington critical luminosity}
\label{ssec_ledd}
We {also} check whether the most
massive models may become gravitationally unbound, i.e. 
 the rate of energy outflow  at the surface 
exceeds the corresponding Eddington luminosity
\begin{equation}
L_{\rm E} = \frac{4 \pi G c M}{\kappa}
\label{eq_edd}
\end{equation} 
where $M$ is the stellar mass, 
and $\kappa$ the opacity of the surface layer.

\begin{figure}
\resizebox{\hsize}{!}{\includegraphics{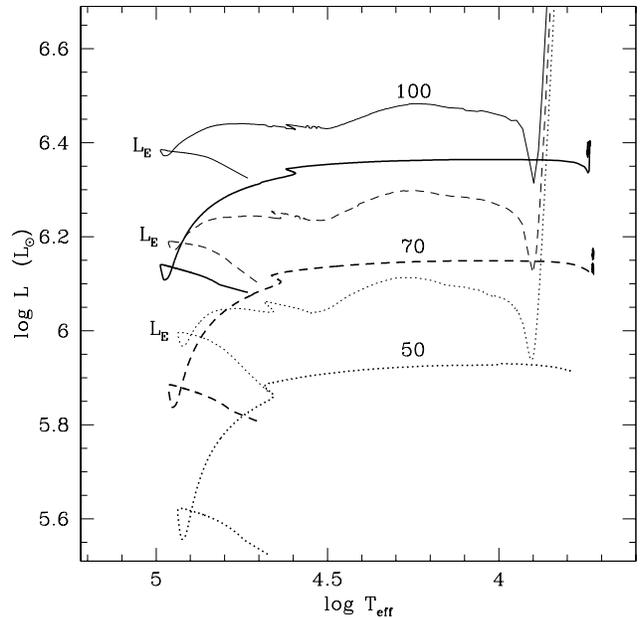}}
\caption{Evolutionary tracks in the H-R diagram (thick lines) for selected
massive models with initial masses  (in $M_{\odot}$) as indicated.
Thin-line curves labelled with $L_{\rm E}$ show the behaviour of the
Eddington luminosity calculated at the photospheric layer with 
Eq.~(\protect\ref{eq_edd}).}
\label{fig_ledd}
\end{figure}

In Fig.~\ref{fig_ledd} we compare the evolutionary paths in the H-R diagram
for selected models of different initial masses with the 
loci obtained  from  Eq.~(\ref{eq_edd}), i.e.
calculating  the Eddington luminosity of the photosphere for each
value of the effective temperature along the evolutionary sequence.

%

\begin{figure}
\resizebox{\hsize}{!}{\includegraphics{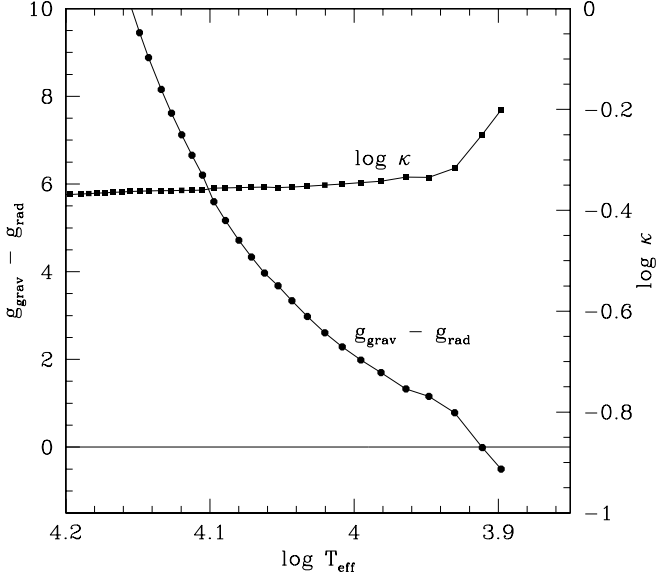}}
\caption{Effective acceleration at the photosphere, defined as
the difference between the inward gravitational acceleration, $g_{\rm grav}$,
and the outward radiative acceleration, $g_{\rm rad}$.
A sequence of models are shown up to the point in which the net acceleration
becomes negative, as the surface luminosity exceeds the Eddington limit.
Correspondingly, the evolution of the photospheric opacity, $\kappa$, 
is shown.}
\label{fig_gravrad}
\end{figure}


From this simple test, it turns out that, 
{in contrast with} the  $50 \, M_{\odot}$ model, 
the $70 \, M_{\odot}$ and $100 \, M_{\odot}$ models
may achieve super-Eddington luminosities 
(typically at $\log T_{\rm eff} \sim 3.9$)
on their way to the Hayashi line,  
towards the end of the He-burning phase. Correspondingly,
the effective acceleration\footnote{The effective acceleration is 
defined as $G M / R^{2} - (\sigma/c) \,\, \kappa T_{\rm eff}^4$, i.e. the
difference between the gravitational and radiative accelerations.}
at the surface becomes negative, due to
the increase of the surface radiative opacity (see Fig.\ref{fig_gravrad}).

From this point onward the predicted evolution should not be
considered reliable, as the adopted  
assumption of hydrostatic equilibrium does not hold any longer for
the outermost layers. Likely, these stars would
start losing mass from their surface.

For the sake of simplicity,  we do not attempt
to include any prescription for mass-loss driven by 
super-Eddington luminosities in the most massive stars in this work.
This, and other stellar winds driving mechanisms, 
will be analysed in more detail in a future {paper} 
dedicated to the evolution of zero-metallicity stars with mass loss 
(Marigo et al. 2001, in preparation).

\section{Surface chemical changes}
\label{sec_chem}
%

\begin{figure}
\resizebox{\hsize}{!}{\includegraphics{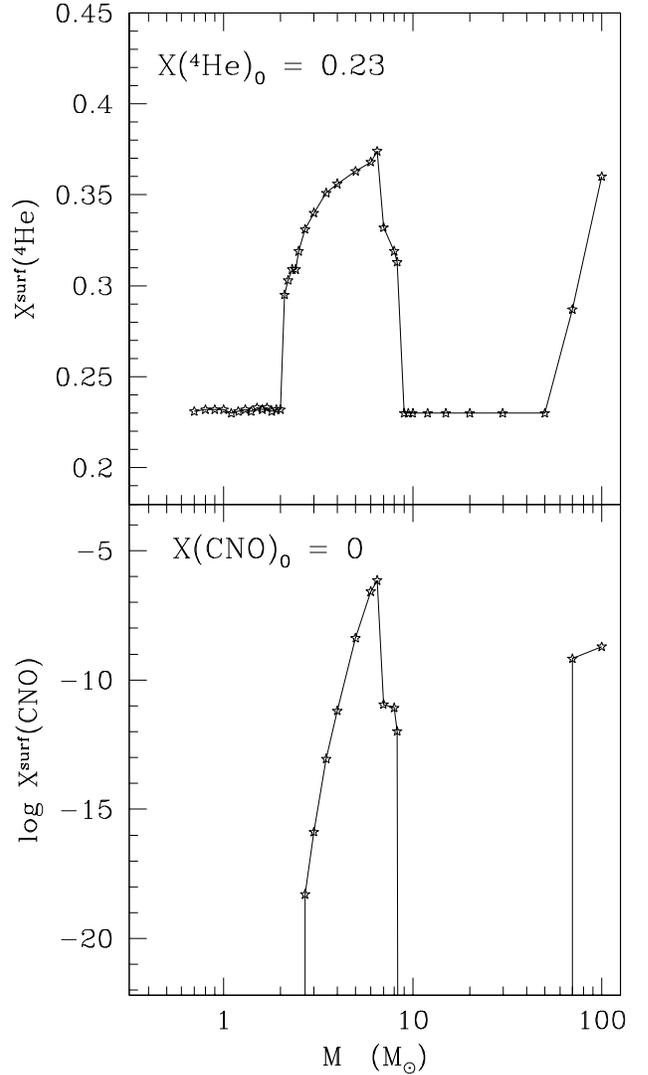}}
\caption{Surface abundances of $^{4}$He and CNO elements as a function
of the stellar mass, for the last computed model.}
\label{fig_chem}
\end{figure}

We will {now} discuss the changes in the surface chemical composition
due to convective dredge-up episodes. Usually these 
result from the penetration of the convective
envelope into stellar layers which have previously been modified
by nucleosynthetic processes.
In this respect it is relevant to recall that the deepest inward extension
of envelope convection occurs as a star reaches close to its Hayashi line.
 Keeping this in mind and looking Table~\ref{tab_du} we can make
the following considerations.

Between the end of the H-burning phase and the central He-ignition, 
the only models that evolve to the red part 
of the H-R diagram and settle onto the giant branches 
(closely approximating the location of the Hayashi tracks) are those
of low-mass ($0.7 \, M_{\odot} \la M \la 1.1 \, M_{\odot}$).
At this stage envelope convection starts to proceed inward, but
it does not extend {significanlty} into the chemical profile
left by the radiative burning core. 
As can be noticed 
in Table~\ref{tab_du}, this {\sl first dredge-up} episode in low-mass stars
changes the surface chemical abundance of helium
by negligible amounts.    
For more massive models ($M > 1.2\, M_{\odot}$) 
the onset of the He-burning phase takes place far away from 
the Hayashi tracks. As a consequence, 
the first notable feature of zero-metallicity models
is that the first dredge-up {essentially} does not take place.

As far as the {\sl second dredge-up} is concerned, we can notice from 
Fig.~\ref{fig_hhe} that 
after central helium exhaustion the evolution 
for all models with $M < 70 \, M_{\odot}$
is characterised  by a redward excursion in the H-R diagram.
However, the minimum value of the effective temperature attained --
either at the start of the AGB phase or at central carbon ignition --   
has a non-monotonic dependence on the stellar mass. 
Specifically, stars with initial masses 
$0.7\, M_{\odot} \la M \la 8\, M_{\odot}$ are able to 
approach their Hayashi tracks, then becoming red giants.
In this mass range the second dredge-up is found to 
actually occur in models with $2.1\, M_{\odot} \la M \la 8.3 \, M_{\odot}$,
the envelope penetration being insufficient at lower masses.
Moreover, we can notice from Table~\ref{tab_du} that for   
$2.1\, M_{\odot} \la M \la 2.7\, M_{\odot}$ only helium increases 
its abundance in the envelope, whereas for 
$2.7\, M_{\odot} \la M \la 8.3\, M_{\odot}$ some primary CNO 
is {also} dredged-up to the surface.   

It is worth remarking that further changes in the surface chemical
abundances may be caused by the occurrence of the {\sl third dredge-up}
at thermal pulses 
(if they take place, see Sect.~\ref{ssec_agb}) during the TP-AGB phase
(see, for instance, Dom\'\i nguez et al. 1999).
Actually, in our calculations we do not find any evidence of this process,
but this is more likely due to the small number of He-shell flashes 
{that have been} followed for each evolutionary sequence.  

More massive stars ($8.3 \, M_{\odot} < M \la 50\, M_{\odot}$) 
met the necessary conditions for carbon ignition in the core
at higher effective temperatures, i.e. 
before entering the giant branch phase. As expected, these 
models do not suffer any dredge-up episode, i.e. their surface
layers preserve the original composition. 

Finally, stars with $70\, M_{\odot} < M \la 100 \, M_{\odot}$
settle on their Hayashi tracks already during the core He-burning
phase, and there they remain until the onset of carbon burning.
The second dredge-up takes place enriching 
the surface composition both in helium and CNO elements
(see Table \ref{tab_du}).
As already discussed in Sect.~\ref{ssec_ledd}, 
the results for these massive and luminous models --  
made to evolve at constant mass -- should be taken with some caution.
Our analysis has in fact revealed that these models may become 
dynamically unstable against 
radiation pressure towards the end of core helium burning.
From that point, stellar winds driven by super-Eddington luminosities
will indirectly affect the surface chemical composition by
i) reducing the stellar mass, and ii) influencing the efficiency 
of the second dredge-up. To this regard, 
a more careful study is going to be carried 
out in a forthcoming work (Marigo et al. 2001, in preparation).

In conclusion, owing to the 
particular evolutionary features of 
the zero-metallicity stars, we find 
that significant changes in the chemical
composition of the envelope - during the evolutionary phases 
here considered -- may occur within two well defined mass ranges, i.e.
$2.1\, M_{\odot} < M \la 8\, M_{\odot}$, and
$70\, M_{\odot} < M \la 100\, M_{\odot}$.
Our results are fully illustrated in Fig.~\ref{fig_chem}.

\section{Might we ever detect primordial stars ?}
\label{sec_strip}
Zero-metallicity stars with initial masses $M < 0.8 \, M_{\odot}$,
if ever born, should be still alive, as their nuclear lifetimes are
longer than the age of the universe (see Table \ref{tab_time}).
However, to date the hypothesis  
of the possible formation of primordial low-mass stars has not been 
{proven observationally}.  
In order to devise a test for the existence of
such stars, one should single out observable 
distinctive features which can point unambiguously to 
their existence.  For instance,  
the absence of metal lines in the stellar spectra would immediately
give us the confirmation, but this circumstance is hard to {meet}.
In fact, the interaction of long-lived first stars 
with a chemically enriched interstellar medium would have probably
polluted their primordial surface compositions 
with newly synthesized metals.

Another peculiar feature of low-mass zero-metallicity stars 
(with $ M \sim 1 \, M_{\odot}$) is the development of a loops in 
the H-R diagram, corresponding to the first activation of the CNO cycle 
as soon as some primary $^{12}$C is produced by 
the 3-$\alpha$ reaction (see Sect.~\ref{ssec_3alpha}).
Unfortunately, the probability of detecting such loop 
in a primordial simple stellar population is negligible,
as this feature is quite short-lived (a few $10^{6}$ yr),
hence scarcely populated.

Finally, distinctive observational properties of
low-mass zero-metallicity stars may {be derived} from
their {loci} in the HR-diagram: they are 
slightly hotter than the Pop-II stars of even the lowest
metallicity.
For instance, Cassisi et al.\ (1996) find that most of
the evolution of $Z=10^{-10}$ HB stars occurs to the left
(higher $T_{\rm eff}$) 
of the RR Lyrae instability strip, {thus} implying a very
low probability of finding RR Lyrae pulsators among an hypothetical
population of these stars. 

However, we notice that high effective temperatures are also met 
at the turn-offs of low-mass stars. This might open the possibility 
that, at ages $>10$ Gyr, the main-sequence of $Z=0$ populations
falls inside the faintest part of the instability strip 
which contains (in order of decreasing luminosity)
three classes of variables, namely:   
Cepheids, RR Lyrae and $\delta$~Scuti stars. In such a case, the possibility 
arises that a large number of short-period, low-amplitude pulsators
might be present among Pop-III stars.
  
\begin{figure}
\resizebox{\hsize}{!}{\includegraphics{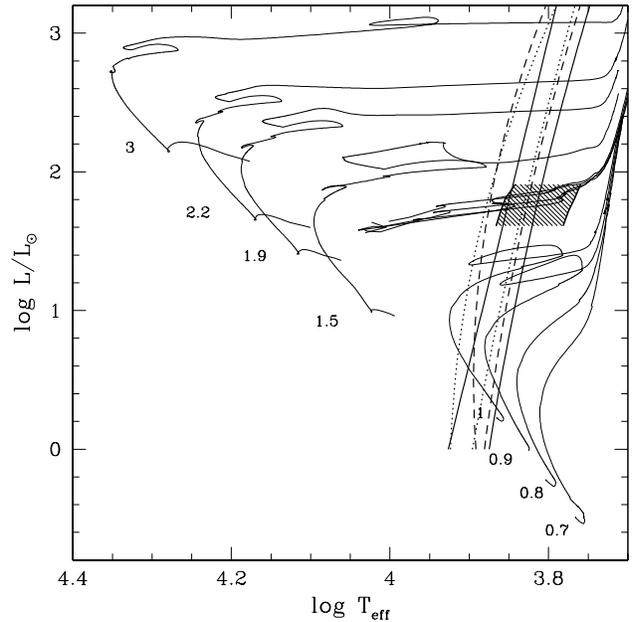}}
\caption{Instability strips in the $\log L - \log T_{\rm eff}$ diagram.
Theoretical edges for ($Z=0$, $Y=0.23$) are drawn assuming pulsation 
occurs in the fundamental mode (solid line), first overtone (dashed line),
and second overtone (dotted line).
The shaded area shows 
the expected location of RR Lyrae 
according to Bono et al. (1995). 
See text for more details.}
\label{fig_instrip}
\end{figure}

\begin{figure}
\resizebox{\hsize}{!}{\includegraphics{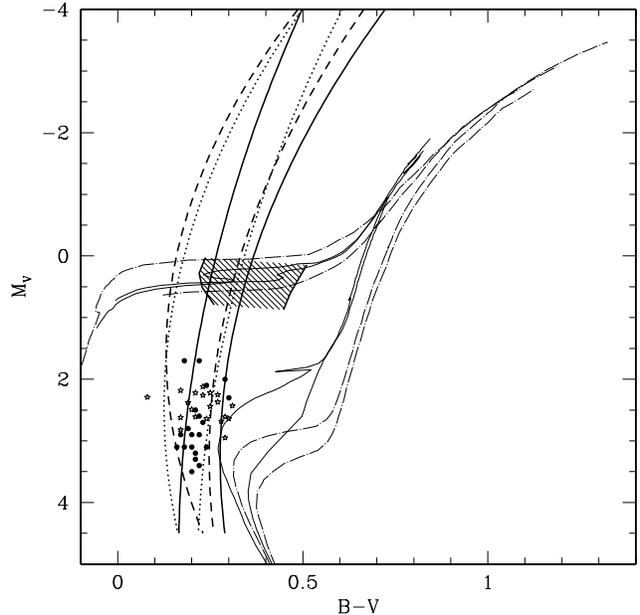}}
\caption{Instability strips in the $M_{V} - (B-V)$ diagram,
with the same notation as in Fig.~\protect\ref{fig_instrip}.
Isochrones of ages 15.8 Gyr and 10.0 Gyr are shown 
for $Z=0$ (continuous line) and $Z=10^{-4}$ (from 
Girardi et al.\ 1996; dot-dashed line).
Points correspond to observed $\delta$~Scuti stars in various
galactic globular clusters and in the Carina DSph galaxy 
(Nemec \& Mateo 1990, filled cirles; McNamara 2000, starred symbols).}
\label{fig_instbv}
\end{figure}

To investigate this possibility, the results of a preliminary analysis 
of pulsation and evolution of Z = 0 stars are shown in 
Figs.~\ref{fig_instrip} and \ref{fig_instbv}.
We have made use of the large data set of theoretical models of Cepheid
variables calculated by Chiosi et al. (1993), corresponding to luminosities
$\log L/L_{\odot} > 2$, various values of the stellar mass
($ 3 M_{\odot} \la M \la 12 M_{\odot}$),  and different metal and helium
abundances ($Z\,=\,$0.02, 0.008, 0.004; $Y\,=\,$0.23, 0.24). 
This data set has been extended towards lower luminosities
by means of a preliminary stability analysis, specifically performed 
for zero-metallicity models with $M = 1 M_{\odot}$.
The linear nonadiabatic pulsation calculations have been carried out
with the aid of the same code as in 
Chiosi et al. (1993; and references therein), but incorporating the 
low temperature opacities of Alexander and Ferguson (1994).
The new calculations cover the region of the HR diagram
delimited by $0.3 \le \log(L/L_\odot) \le 1.3$ and 
$3.90 \ga \log T_{\rm eff} \ga 3.73$. 

Combining the earlier pulsation models with the
new ones, we are now able to determine 
the edges of the instability strip 
from $\log L/L_{\odot} \sim 4.5$  down to $\log L/L_{\odot} \sim 0$,
as illustrated in  Fig.~\ref{fig_instrip}. 
Analogously to Chiosi et al. (1993), we have also derived several 
parameterised fitting formulas (given in Table~\ref{tab_fitting})
as a function of helium $Y$ and metal content $Z$. 
They describe, for different pulsation modes, the location of the instability
strips and other relevant  relations between period, luminosity, colour, 
and mass. The instability strips displayed in Fig.~\ref{fig_instrip}
are obtained from the aforementioned analytical relations 
for the red and blue edge setting $Z=0$ and $Y=0.23$. 

The main interesting points are the following.
At luminosities $\log L/L_{\odot} > 2$ the instability strip 
crosses the evolutionary tracks of primordial stars which should 
already be dead and hence undetectable.  
Towards lower luminosities the 
theoretical location of the instability strip 
is consistent with the predictions  
for RR Lyrae variables according to Bono et al. (1995)
for fixed  helium and metal content ($Y=0.24$, $Z=0.001$).
The shaded region corresponds 
to the maximum extension of the strip (to the red and to the blue), 
combining the results of Bono et al. (1995) 
for both fundamental and first overtone pulsators.
We can notice that the RR Lyrae instability strip is almost 
twice as extended to the red as our instability strip 
at the same luminosities.
It is worth remarking here that, 
whereas the blue edge is well determined by our stability analysis, 
the red edge is not (for more details see Chiosi et al. 1993). 
As a conservative approach, we define the 
red edge to coincide with the maximum of the growth {rate}. 
In this way, we are certainly underestimating the width of the 
instability strip. 

Anyhow, at the typical luminosities of RR Lyrae, the instability
strip for the $Z=0$ models would cover the red portions of the HBs 
described by the zero-metallicity low-mass 
tracks plotted in  Fig.~\ref{fig_instrip} (see Cassisi et al.
1996, for a similar result).
In other words, pulsation should set in during a period of 
very rapid evolution, i.e. only after these stars have
burnt most of their central helium in the bluer regions close
to the ZAHB. This again would make the observability almost out
of reach.

Finally, going to even fainter luminosities ($\log L/L_{\odot} \la 1$) 
it turns out that zero-metallicity
models with mass as low as $0.9 \, M_\odot$ are expected to exhibit
pulsational properties during the main-sequence phase. 
At the luminosities and effective temperatures under consideration
(see Fig.~\ref{fig_instrip}), 
the pulsation periods can range from about 0.6 to 0.03 days.
The reliability of these provisional results is 
supported to some extent by the fact that the location 
on the H-R diagram and the period determinations
are consistent with the observational data of $\delta$~Scuti stars -- 
a rich class of faint variables with
amplitudes ranging from hardly detectable to several tenths of a
magnitude -- which are usually present among the Blue Stragglers 
of many Galactic globular clusters.
A sample of observed $\delta$~Scuti stars
(Nemec \& Mateo 1990; McNamara 2000) 
is shown in the $M_V$ versus $B-V$ diagram 
of Figure~\ref{fig_instbv}, and compared to the theoretical 
instability strip.
It is interesting to notice that, applying the fundamental-mode
period-mass-luminosity-colour relation presented in Table~\ref{tab_fitting},
to the $\delta$~Scuti data (Nemec \& Mateo 1990; McNamara 2000) 
and assuming $Z=0.001$, 
we can immediately estimate the pulsational 
masses for these variables, with a typical range between 
$1.2$ and  $1.8 \, M_\odot$.

A direct comparison with observations is presented in  
Fig.~\ref{fig_instbv}, showing isochrones representative 
of Pop-II ($Z=10^{-4}$) and Pop-III 
($Z=0$; see the Appendix) stellar populations, for two
values of the age, i.e. 10.0 Gry and 15.8 Gyr.
These latter define the best range for the age of Galactic
globular clusters, as recently reviewed by Carretta et al. (2000). 
 
We can see that in none of the four
cases displayed, the
turn-off region is located inside the instability strip, 
except for the $Z=0$, 10.0 Gyr old isochrone, 
which marginally intersects the red edge of the strip.
However, we note that,
since our calculations probably underestimate the red extension 
of the instability strip 
(see earlier in this section),
there is still some possibility that the $Z=0$ isochrones 
for ages within the range under consideration 
will enter significantly into the red part of the instability strip. 
If this is indeed the case, then it implies that a 
large number of primordial pulsators could be
present in the present era.  Such stars would presumably be
similar to $\delta$~Scuti variables but with lower masses 
at a given colour (since Z = 0 stars near the 
main-sequence are bluer and more luminous than solar metallicity
stars of the same mass).  They would also have longer periods
at a given luminosity and colour (since pulsation period varies
approximately as $M^{-0.5}$). 

{Might we ever detect them ?}
\begin{table*}
\begin{center}
\caption{Fitting relations for pulsational models. Coefficients
are given for the fundamental mode, first and second overtone.
Respective periods ($P_0$, $P_1$, and $P_2$) are expressed in days,
masses and luminosities are in solar units, helium ($Y$) and metal ($Z$)
abundances are by mass. Standard notation of the quantities is used.}
\label{tab_fitting}
\begin{tabular*}{160mm}{l ccc ccc ccc}
\noalign{\smallskip}\hline\noalign{\smallskip}
Mode   &    A     &   B       &   C        &    D        &  E  &F &G &H &I    \\
\noalign{\smallskip}
\hline
\noalign{\smallskip}
\multicolumn{10}{c}
{Blue Edge: log T$_{BE}$ = A + B$\times$Y + C$\times$Z + 
D$\times$(logM) + E$\times$(logL) + F$\times$(logL)$^2$
+ G$\times$Y(logL)$^2$ + H$\times$Z(logL)$^2$}\\
\noalign{\smallskip}
\hline
\noalign{\smallskip}
P$_0$  &   3.928  &  2.985E-2 &  -5.021E-1 &    2.986E-2 &  -5.130E-2
 & -1.311E-3 & 5.915E-3 & -2.319E-2 & \\
P$_1$  &   3.962  &  -2.0117E-1  &  -6.903E-1  &  8.774E-2  &  -1.298E-2  
 & -1.945E-2 & 3.073E-2 & -2.658E-2 & \\
P$_2$  &   3.936  &  4.816E-2  &  -8.542E-1  &  8.414E-2  &  -3.018E-2   
 & -1.515E-2 &  1.758E-2  &   6.488E-2 & \\
\noalign{\smallskip}
\hline
\noalign{\smallskip}
\multicolumn{10}{c}
 {Red  Edge:  logT$_{RE}$ =  A + B$\times$Y + C$\times$Z + 
D$\times$(logM) + E$\times$(logL) + F$\times$(logL)$^2$ 
+ G$\times$Y(logL)$^2$ + H$\times$Z(logL)$^2$} \\
\noalign{\smallskip}
\hline
\noalign{\smallskip}
P$_0$  &  3.879 &  5.948E-2 & 2.274E-1 &  6.708E-2 & -4.646E-2 & -3.762E-3 & 
-4.628E-4 & -1.511E-1 &\\
P$_1$  &   3.901 & -7.233E-3 & 5.884E-2 &  6.731E-2 & -4.071E-2 & -6.632E-3 &  
6.754E-3 & -1.177E-1 &\\
P$_2$  &  3.876 &  1.134E-1 & -5.327E-1 &  2.224E-2 & -3.824E-2 & -3.246E-3 &  
2.557E-3 &  3.747E-2 &\\
\noalign{\smallskip}
\hline
\noalign{\smallskip}
\multicolumn{10}{c} 
{Blue Edge:  (B-V)$_{BE}$ = A + B$\times$Y + C$\times$Z + D$\times$(M$_V$) + E$\times$(Y$\times$M$_V$) 
 + F$\times$(Z$\times$M$_V$) + G$\times$(M$_V^2$) + H$\times$(Y$\times$M$_V^2$) + I$\times$(Z$\times$M$_V^2$)}\\
\noalign{\smallskip}
\hline
\noalign{\smallskip}
P$_0$ &  3.102E-1 & -1.925E-1 &  5.349 &  -5.217E-2 &  4.720E-2 &  -2.590E-1 &  6.541E-3 & -1.019E-2 &  1.516E-1\\
P$_1$ & -2.215E-1 &  1.653E   &  8.027 &  -3.594E-1 &  1.406    &   2.767    & -2.355E-2 &  1.524E-1 &  7.409E-1\\
P$_2$ &  1.062E-1 &  3.175E-1 &  4.482 &  -2.750E-1 &  1.004    &   2.314    & -2.932E-2 &  1.664E-1 &  7.820E-1\\
\noalign{\smallskip}
\hline
\noalign{\smallskip}
\multicolumn{10}{c}
{Blue edge:   (V-I)$_{BE}$ =  A + B$\times$Y + C$\times$Z + D$\times$(M$_V$) + E$\times$(Y$\times$M$_V$) 
+ F$\times$(Z$\times$M$_V$) + G$\times$(M$_V^2$) + H$\times$(Y$\times$M$_V^2$) + I$\times$(Z$\times$M$_V^2$)}\\
\noalign{\smallskip}
\hline
\noalign{\smallskip}
P$_0$ &  5.029E-1 & -4.736E-1 &  2.331 &  -3.438E-2 & -9.280E-2 & -1.857E-1 &  6.263E-3  &-2.051E-2  & 3.047E-2\\
P$_1$ &  8.104E-2 &  8.586E-1 &  3.731 &  -2.736E-1 &  9.622E-1 &  1.408    & -1.644E-2  & 1.071E-1  & 3.687E-1\\
P$_2$ &  3.184E-1 & -1.931E-1 &  1.809 &  -2.387E-1 &  7.655E-1 &  2.243    & -2.424E-2  & 1.377E-1  & 6.635E-1\\
\noalign{\smallskip}
\hline
\noalign{\smallskip}
\multicolumn{10}{c}
{Red edge:  (B-V)$_{RE}$ =  A + B$\times$Y + C$\times$Z + D$\times$(M$_V$) + E$\times$(Y*M$_V$) 
+ F$\times$(Z$\times$M$_V$) + G$\times$(M$_V^2$) + H$\times$(Y$\times$M$_V^2$) + I$\times$(Z$\times$M$_V^2$)}\\
\noalign{\smallskip}
\hline
\noalign{\smallskip}
P$_0$ &  3.349E-1 &  1.407E-1 &  4.187 &  -1.405E-1 &  3.711E-1 & -1.638    & -5.235E-3  & 5.933E-2  & 3.714E-1\\
P$_1$ &  1.341E-1 &  8.584E-1 &  6.649 &  -2.564E-1 &  8.899E-1 &  1.708    & -2.054E-2  & 1.237E-1  & 9.079E-1\\
P$_2$ &  3.258E-1 &  7.263E-2 &  4.307 &  -1.777E-1 &  5.641E-1 &  2.764E-1 & -1.799E-2  & 9.780E-2  & 2.353E-1\\
\noalign{\smallskip}
\hline
\noalign{\smallskip}
\multicolumn{10}{c} 
{Red edge:  (V-I)$_{RE}$ =  A + B$\times$Y + C$\times$Z + D$\times$(M$_V$) + E$\times$(Y*M$_V$) 
+ F$\times$(Z$\times$M$_V$) + G$\times$(M$_V^2$) + H$\times$(Y$\times$M$_V^2$) + I$\times$(Z$\times$M$_V^2$)}\\
\hline
\noalign{\smallskip}
P$_0$ &  5.525E-1 & -1.107E-1 &  3.013E-1 &  -9.441E-2 &  1.655E-1 & -1.415   &  -1.833E-3 &  3.032E-2 &  1.165E-1\\
P$_1$ &  3.827E-1 &  4.601E-1 &  2.334    &  -1.970E-1 &  6.056E-1 &  1.088   &  -1.605E-2 &  8.835E-2 &  5.258E-1\\
P$_2$ &  5.315E-1 & -1.852E-1 &  6.837E-1 &  -1.495E-1 &  4.114E-1 &  1.909E-2&  -1.790E-2 &  8.645E-2 &  2.669E-2\\
\noalign{\smallskip}
\hline
\noalign{\smallskip}
\multicolumn{10}{c}
    { logP = A + B$\times$(logL) + C$\times$(logL)$^2$  + D$\times$logM + E$\times$logT$_{\rm eff}$}\\  
\noalign{\smallskip}
\hline
\noalign{\smallskip}
P$_0$  &  12.150  &  0.659    &  3.000E-2  &  -0.723 &  -3.567 &&&&\\ 
P$_1$  &  10.256  &  0.683    &  2.269E-2  &  -0.671 &  -3.112 &&&&\\ 
P$_2$  &  10.002  &  0.692    &  1.978E-2  &  -0.685 &  -3.070 &&&&\\ 
\noalign{\smallskip}
\hline
\noalign{\smallskip}
\multicolumn{10}{c}
{logP = A$\times$logM + B + C$\times$Z + D$\times$M$_V$ + E$\times$(Z$\times$M$_V$) 
+ F$\times$(B-V) + G$\times$(Z$^2$$\times$(B-V)) + H$\times$(B-V)$^2$ + I$\times$(Z$^2$$\times$(B-V)$^2$)}\\
\noalign{\smallskip}
\hline
\noalign{\smallskip}
P$_0$ & -6.656E-1 & -3.771E-1 & -7.788 & -3.411E-1 & 4.206E-1 & 6.834E-1 
& 5.034E+2 & 1.095E-1 & -2.927E+2\\
P$_1$ & -6.072E-1 & -5.719E-1 & -8.367 & -3.254E-1 & 2.656E-1 & 8.998E-1 
& 4.089E+2 & -7.956E-2& -1.571E+2\\
P$_2$ & -5.760E-1 &  -7.067E-1&  -1.016E+1&  -3.099E-1&  -4.138E-1&   1.035    
&  1.092E+3&  -1.451E-1&  -1.345E+3\\
\noalign{\smallskip}
\hline
\noalign{\smallskip}
\multicolumn{10}{c}
{logP = A$\times$logM + B + C$\times$Z + D$\times$M$_V$ + E$\times$(Z$\times$M$_V$) 
+ F$\times$(V-I) + G$\times$(Z$^2$$\times$(V-I)) + 
H$\times$(V-I)$^2$ + I$\times$(Z$^2$$\times$(V-I)$^2$)}\\
\noalign{\smallskip}
\hline
\noalign{\smallskip}
P$_0$ & -6.864E-1 & -3.636E-1 & -2.742 & -3.347E-1 & -8.353E-1 & 2.243E-1 
& 1.317E+2 & 5.397E-1 & -1.859E+2\\
P$_1$ & -6.103E-1 & -5.764E-1 & -5.814 & -3.159E-1 & -1.017    & 4.879E-1 
& 2.659E+2 & 3.382E-1 & -2.217E+2\\
P$_2$ & -5.661E-1 & -7.197E-1 & -6.262 & -3.007E-1 & -1.382    &   6.468E-1
&  6.420E+2& 1.995E-1 & -8.657E+2\\
\noalign{\smallskip}\hline\noalign{\smallskip}
\end{tabular*}
\end{center}
\end{table*}

\begin{acknowledgements}
It is a pleasure to thank 
B. Salasnich for his help with the evolutionary code,
and A. Weiss for useful conversations on Pop-III stars.
A sincere thank goes to the referee whose comments contributed to 
improve the presentation of the work.
This study is funded by the Italian Ministry of University, Scientific 
Research and Technology (MURST) under contract ``Formation and evolution
of Galaxies'' n.~9802192401.
\end{acknowledgements} 

{}

\appendix

\section{Tables of evolutionary tracks}
\label{sec_tabletrack}

The data tables for the present evolutionary tracks are available only in
electronic format, either upon request to the authors, or {by } accessing  
a WWW site containing a complete data-base (including additional data 
and the future extensions) at http://pleiadi.pd.astro.it. 

For each evolutionary track, the corresponding data file is organised  
into \ref{item_stage} columns with the following information:
	\begin{enumerate}
	\item \verb$age/yr$: stellar age in yr;
	\item \verb$logL$: logarithm of surface luminosity (in solar units), 
\logL;
	\item \verb$logTef$: logarithm of effective temperature (in K), 
\logTe;
	\item \verb$grav$: logarithm of surface gravity ( in cgs units);
	\item \verb$logTc$: logarithm of central temperature (in K);
	\item \verb$logrho$: logarithm of central density (in cgs units);
	\item \verb$Xc,Yc$: mass fraction of either hydrogen (up to the 
central H-exhaustion) or helium (later stages) in the stellar centre;
	\item \verb$Xc_C$: mass fraction of carbon in the stellar centre;
	\item \verb$Xc_O$: mass fraction of oxygen in the stellar centre;
	\item \verb$Q_conv$: fractional mass of the convective core;
	\item \verb$Q_disc$: fractional mass of the first mesh point where 
the chemical composition differs from the surface value;
	\item \verb$L_H/L$: the luminosity 
provided by H-burning reactions {as a fraction of} the surface luminosity;
	\item \verb$Q1_H$: fractional mass of the inner border of the 
H-rich region;
	\item \verb$Q2_H$: fractional mass of the outer border of the 
H-burning region;
	\item \verb$L_He/L$: the luminosity 
provided by He-burning reactions {as a fraction of} 
 the surface luminosity;
	\item \verb$Q1_He$: fractional mass of the inner border of the 
He-burning region;
	\item \verb$Q2_He$: fractional mass of the outer border of the 
He-burning region;
	\item \verb$L_C/L$: the luminosity 
provided by C-burning reactions {as a fraction of} the surface luminosity;
	\item \verb$L_nu/L$: the luminosity 
{\em lost} by neutrinos (hence negative) {as a fraction of} 
the surface luminosity;
	\item \verb$Q_Tmax$: fractional mass of the point with the highest 
temperature inside the star;
	\item \verb$stage$: label indicating particular evolutionary stages.
\label{item_stage}
	\end{enumerate}

A number of evolutionary stages are 
indicated along the tracks (column~\ref{item_stage}), namely: 
the initial evolutionary stages (\verb$ZAMS$ or 
\verb$ZAHB$), local maxima and minima of $L$ and \Teff\ (\verb$Te-M$, 
\verb$Te-m$, \verb$L-M$, and \verb$L-m$), the exhaustion of central 
hydrogen (\verb$Xc=0$) and helium (\verb$Yc=0$), the 
lowest $L$ and highest \Teff\ during the He-burning of 
intermediate-mass stars (\verb$Bhe$ and \verb$LpT$, respectively), 
the base and tip of the first ascent of the red giant branch (\verb$Brg$ 
and \verb$Tip$, respectively), the maximum $L$ immediately preceding a 
thermal pulse (\verb$1tp$), and the onset of C-burning (\verb$Cb$).
These stages delimit characteristic evolutionary phases, and can be useful 
for the derivation of physical quantities (as e.g.\ typical lifetimes) as a 
function of either mass or metallicity.
Notice that some of these evolutionary stages may be absent from
particular tracks, depending on the precise value of stellar mass 
and metallicity.

\section{Tables of isochrones}
\label{sec_tableisoc}

\begin{figure}
\resizebox{\hsize}{!}{\includegraphics{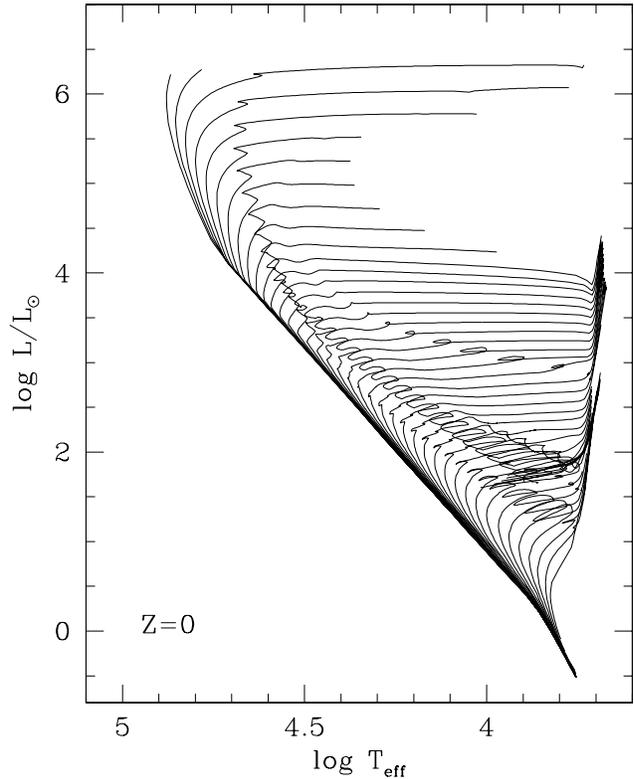}}
\caption{  
Theoretical isochrones in the HR diagram for the initial 
composition $[Z=0, Y=0.23]$. The age range 
goes from $\logt=6.3$ to 10.2, at equally spaced intervals of
$\Delta\log t=0.3$. In all isochrones, the main sequence is
complete down to 0.7~\Msun.}
\label{fig_isocrone}
\end{figure}

From the tracks presented in this paper, we have constructed isochrones
adopting the same algorithm of ``equivalent evolutionary points'' 
as used in Bertelli et al.\ (1994) and Girardi et al.\ (2000). 
The initial point of each isochrone is a 0.7~\Msun\ model in the 
main sequence. The terminal stage of the isochrones corresponds either to 
stars at the beginning of the TP-AGB for ages $\ga10^8$~yr, 
or to stars at C-ignition for $\la10^8$~yr.
The youngest isochrones (up to 6~Myr) are basically limited to the
main sequence stages, since at these ages the inclusion of more 
evolved stars would require tracks for $M\ga100$~\Msun, that will
be presented in a forthcoming paper. 
                       
Complete tables with the isochrones can be obtained 
upon request to the authors, or through the WWW site 
http://pleiadi.pd.astro.it. 
They cover the complete age range from about 2~Myr to 16~Gyr
($6.3<\log(t/{\rm yr})<10.25$). Isochrones are provided
at $\Delta\log t=0.05$ intervals; this means that any two consecutive
isochrones differ by only 12 percent in their ages.

Figure~\ref{fig_isocrone} shows some of these isochrones 
on the HRD, sampled at age intervals of $\Delta\log t=0.3$. 

Theoretical luminosities and effective temperatures along the isochrones 
are translated to magnitudes and colors by using extensive tabulations of 
bolometric corrections and colors, as in Bertelli et al.\ (1994). 
The tabulations were obtained from convolving the spectral 
energy distributions contained in the $\rm [M/H]=-2.0$ 
library of stellar spectra of 
Kurucz (1992) with the response function of several broad-band filters. 
The response functions are from Buser \& Kurucz 
(1978) for the $UBV$ pass-bands, from Bessell (1990) for the $R$ and $I$ 
Cousins, and finally from Bessell \& Brett (1988) for the $JHK$ ones.

For each isochrone table, the layout is as follows: 
A header presents the basic information about the age and metallicity
of each isochrone. Column~1 presents the logarithm of the age in yr;
columns~2 and 3 the initial and actual stellar masses, 
respectively. We recall that the initial mass is the useful quantity
for population synthesis calculations, since together with the initial 
mass function it determines the relative number of stars in different
sections of the isochrones. Then follow the logarithms of surface 
luminosity (column 4), effective temperature (column 5), and surface 
gravity (column 6). From columns 7 to 15, we have the sequence of 
absolute magnitudes, starting with the bolometric one and following 
those in the $UBVRIJHK$ pass-bands. In the last column (16), the 
indefinite integral over the initial mass $M$ of the initial mass 
function (IMF) by number, i.e.\ 
	\begin{equation}
\mbox{\sc flum} = \int\phi(M) \diff M
	\end{equation}
is presented, for the case of the Salpeter IMF, $\phi(M)=AM^{-\alpha}$, 
with $\alpha=2.35$. When we assume a normalization constant of $A=1$, 
{\sc flum} is simply given by {\sc flum}$ = M^{1-\alpha}/(1-\alpha)$.
This is a useful quantity since the difference between any two 
values of {\sc flum} is proportional to the number of stars located in 
the corresponding mass interval. It is worth remarking that  
{\sc flum} relations can be derived
for alternative choices of the IMF, by using 
the values of the initial mass we present in the column 2 of the
isochrone tables.

We also provide summary tables containing
basic information for the most significant stages along the 
isochrones. The evolutionary stages listed are, in sequence:
	\begin{itemize}
	\item \verb$TO$: the turn-off point, i.e.\ 
the point of highest \Teff\ during the core-H burning phase.
	\item If present, \verb$Te-m$ and \verb$Te-M$ signal 
the coldest and hottest points, respectively, of stars in the
overall contraction phase at the end of core$-$H burning. 
	\item \verb$RGBb$ and \verb$RGBt$: the base and tip of the RGB,
	respectively.
	\item \verb$BHeb$: the beginning of the CHeB phase. 
	\item If present, \verb$Te-m$ and \verb$Te-M$ signal 
the coldest and hottest points, respectively, for CHeB stars.
For the youngest isochrones, \verb$Te-M$ represents the maximum 
extension of the Cepheid loop.
	\item \verb$EHeb$: the end of the CHeB phase.
	\item In the oldest isochrones, \verb$L-M$ and \verb$L-m$ 
limit the luminosity range of early-AGB stars; this interval 
corresponds to the clump of early-AGB stars in colour-magnitude 
diagrams.
	\item \verb$1TP$: the beginning of the thermally pulsing 
AGB phase. 
	\item \verb$Cb$: the stage of C-ignition in the cases the 
AGB phase do not occur.
	\end{itemize}

In addition, we provide tables with the integrated broad-band colours 
of single-burst stellar populations. Such tables are derived
by integrating the stellar luminosities, weighted by the predicted
number of stars in each bin, along the isochrones.

\end{document}